\documentclass[conference]{IEEEtran}
\IEEEoverridecommandlockouts
\usepackage{cite}
\usepackage{amsmath,amssymb,amsfonts}
\usepackage{algorithmic}
\usepackage{textcomp}
\usepackage{xcolor}
\usepackage{times}
\usepackage{dsfont}
\usepackage{epsfig,verbatim}
\usepackage{subfigure}
\usepackage{setspace}
\usepackage{color}
\usepackage{cite}
\usepackage{epstopdf}
\usepackage{graphics}
\usepackage{accents}
\usepackage{acronym,mathrsfs}
\usepackage[bookmarks,colorlinks]{hyperref}
\usepackage{booktabs, tabularx}
\usepackage{mathtools}
\usepackage{algorithm}
\usepackage{multirow}
\usepackage{makecell} 

\usepackage{enumitem}
\usepackage{bbm}
\def\BibTeX{{\rm B\kern-.05em{\sc i\kern-.025em b}\kern-.08em
    T\kern-.1667em\lower.7ex\hbox{E}\kern-.125emX}}

\acrodef{dma}[DMA]{dynamic metasurface antenna}
\acrodef{snr}[SNR]{signal-to-noise ratio}
\acrodef{sinr}[SINR]{signal-to-interference-and-noise ratio}
\acrodef{bs}[BS]{base station} 
\acrodef{em}[EM]{electromagnetic} 
\acrodef{mimo}[MIMO]{multiple-input multiple-output}
\acrodef{ris}[RIS]{reconfigurable intelligent surface}
\acrodef{his}[HIS]{holographic intelligent surface}
\acrodef{awgn}[AWGN]{additive white Gaussian noise} 
\acrodef{ula}[ULA]{uniform linear array}
\acrodef{upa}[UPA]{uniform planar array}
\acrodef{isac}[isac]{dual-function radar-communication}
\acrodef{his}[his]{hybrid \ac{ris}}
\acrodef{fgs}[FGS]{fast grid search}
\acrodef{agd}[AGD]{auto gradient descent}
\acrodef{rf}[RF]{radio frequency}
\acrodef{fov}[FOV]{field of view}
\acrodef{ga}[GA]{genetic algorithm}
\acrodef{sdp}[SDP]{semidefinite programming}
\acrodef{sdr}[SDR]{semidefinite relaxation}
\acrodef{isac}[ISAC]{integrated sensing and communication}
\acrodef{mdd}[mDD]{micro-deformation displacement}
\acrodef{md}[m-D]{micro-deformation}
\acrodef{mdm}[mDM]{micro-deformation monitoring}
\acrodef{tdd}[TDD]{time-division duplex}
\acrodef{sar}[SAR]{synthetic aperture radar}
\acrodef{ltm}[LTM]{learnable template-matching}
\acrodef{ai}[AI]{artificial intelligence}
\acrodef{ofdm}[OFDM]{orthogonal frequency-division multiplexing}
\acrodef{sar}[SAR]{synthetic aperture radar}
\acrodef{cnn}[CNN]{convolution neural network}
\acrodef{fft}[FFT]{fast Fourier transform}

\begin{document}

\title{Learnable Template Matching Approach for Micro-Deformation Monitoring based on Integrated Sensing and Communication Platform
}

\author{  
	\IEEEauthorblockN{Zhuoyang Liu,~\IEEEmembership{Student Member,~IEEE}, Yixiang Luomei,~\IEEEmembership{Member,~IEEE}, Feng Xu,~\IEEEmembership{Senior Member,~IEEE}\\
	} 
%  \IEEEauthorblockA{{$^*$}
%  \textit{Key Lab for Information Science of Electromagnetic Wave (MoE)}, 
% \textit{Fudan University},
% Shanghai 200433, China,\\}
%  \IEEEauthorblockA{{$^\dagger$}
%  \textit{National Key Laboratory of Science and Technology on Communications},
% \textit{University of Electronic Science and Technology of China},
% Sichuan 100084, China,\\}
%  \IEEEauthorblockA{{$^\ddagger$}
%  \textit{School of Communication and Information Engineering},
% \textit{Nanjing University of Posts and Telecommunications},\\
% Nanjing 210003, China,\\}
% \IEEEauthorblockA{{$^+$}
%  \textit{Faculty of Math and CS},
% \textit{Weizmann Institute of Science},
% Rehovot, Israel,\\}
% \IEEEauthorblockA{Email:liuzy20@fudan.edu.cn.}
\thanks{
    Z. Liu, Y. Luomei and F. Xu are with the Key Lab for Information Science of Electromagnetic Wave (MoE), Fudan University, Shanghai 200433, China (e-mail: \{liuzy20; lmyx; fengxu\}@fudan.edu.cn).
    
    % Y. Zhang is with the National Key Laboratory of Science and Technology on Communications, University of Electronic Science and Technology of China, Sichuan, China.
    
    % H. Zhang is with the School of Communication and Information Engineering, Nanjing University of Posts and Telecommunications, Nanjing 210003, China.
    
    % Y. C. Eldar is with the Faculty of Math and CS, Weizmann Institute of Science, Rehovot, Israel (e-mail: yonina.eldar@weizmann.ac.il).
    }
	
	\vspace{-1.0cm}

 }

\maketitle

\begin{abstract}
    Existing \ac{isac} platforms fail to fully utilize the shared spectrum and aperture resources for sensing, resulting in poor sensing performance. Specifically, weak target sensing on the \ac{isac} platform, such as micro-deformation monitoring (mDM), suffers from inaccurate measurements due to poor sensing quality. In this paper, we propose an AI-assisted approach to alleviate the effect of poor sensing quality in the \ac{isac} system by effectively removing the clutter. We begin by modeling the environment clutter model as a combination of the deterministic and stochastic signals to represent urban coverage scenarios around the \ac{bs}. A clutter suppression optimization problem is formulated to extract the micro-deformation displacement (mDD) from the original \ac{isac} signals. We then propose a learnable template-matching (LTM) approach to mitigate the influences of clutters, thereby enhancing sensing quality. In particular, the \ac{em} signal feature of the mDD is embedded into the network to strengthen the mDM signal, and clutter filters are incorporated to suppress environmental clutter. Numerical results illustrate the superiority of our proposed approach concerning convergence speed and accuracy in mDD prediction. By deploying our approach to the \ac{bs} measurement, the simulation-only trained LTM exhibits impressive performance in environment clutter separation and mDD estimation.

\end{abstract}


\section{Introduction}
\label{intro}

\begin{figure}[htbp]
    \centering
    \includegraphics[width=.95\linewidth]{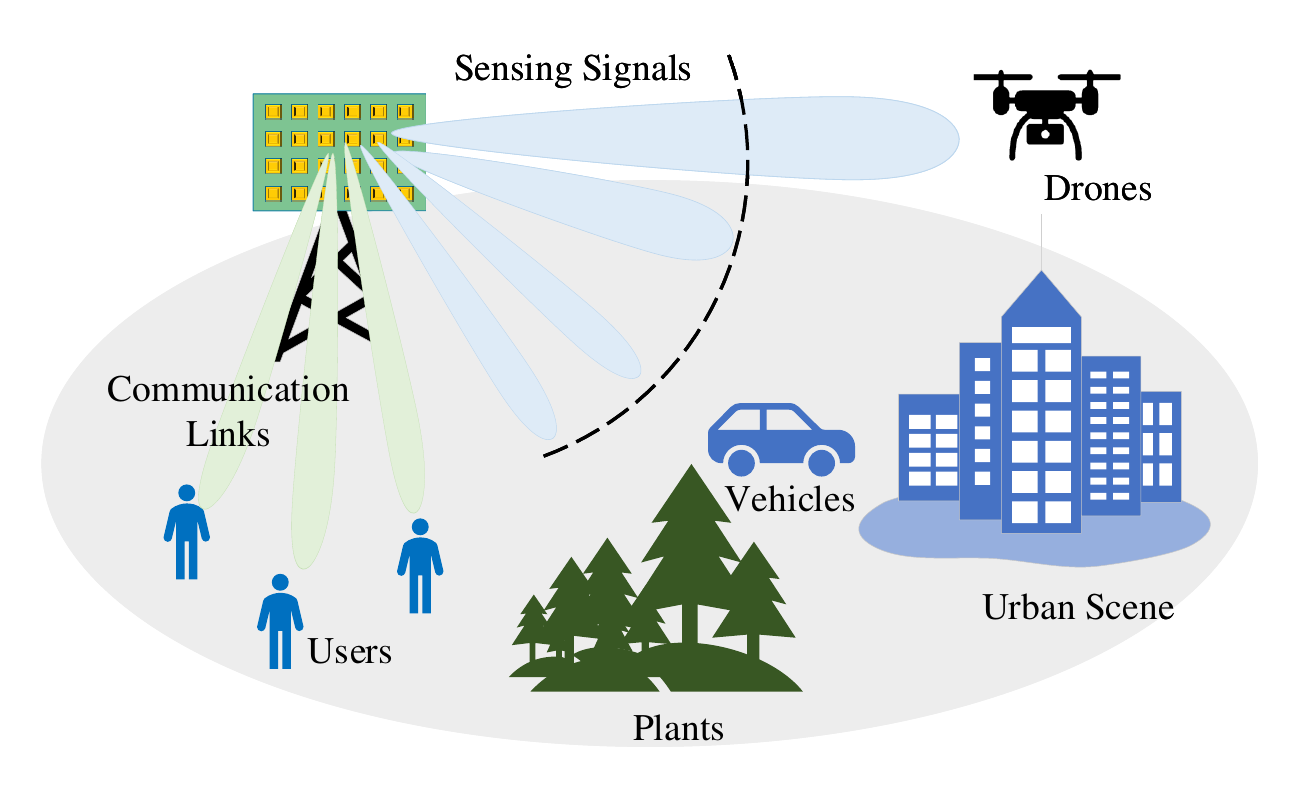}
    \caption{ISAC application in smart city.}
    \label{fig:isac}
\end{figure}

The \acf{isac} system is considered an essential transition from 5G to 6G, effectively tackling the spectrum and aperture congestion by enabling the simultaneous execution of wireless communication and sensing on a unified platform \cite{6G1,6G3}.
\ac{isac} systems play crucial roles in many applications, including but not limited to autonomous driving \cite{yang2021edge}, smart city \cite{6G4}, perimeter security \cite{wu2024ai}, and infrastructure monitoring \cite{cheng2022integrated}.
These applications, as shown in Fig.~\ref{fig:isac}, rely on high-speed and reliable communication connectivity, as well as high-precision and exceptional sensing quality.
To achieve a balance between communication and sensing performance in \ac{isac} systems, many researchers have focused on exploring effective resource management strategies. 
However, compared to a dedicated radar system, the \ac{isac} system is constrained in its capability to allocate resources for sensing functionality, leading to inadequate sensing performance.

{\color{black}Our research aims to tackle weak target sensing problems based on the \ac{isac} platforms, with a particular focus on \ac{mdm}, striving to overcome measurement limitations related to low sensing quality.
Specifically, the \ac{md} refers to a small structural displacement occurring within a single range bin and significantly smaller than the structural dimensions and sensing distance.
Thus, the \ac{mdm} task is a typical weak target detection problem and plays a vital role in ensuring the safety of urban transportation networks \cite{zhao2022dynamic,tian2019vibration} and facilitating the effective maintenance of large-scale infrastructure \cite{pramudita2023fmcw,mao2025avisual}.}
Unlike traditional accelerometer-based \cite{lim2020vibration,ma2023continuous} and radar-based \ac{mdm} systems \cite{pramudita2023fmcw,yang2023multi,sadeghi2023using,ma2023structural}, the \ac{isac} platform achieves \ac{mdm} by sequentially performing communication and sensing operations in a \ac{tdd} mode on the \acf{bs}.
With the extensive coverage and continuous transmission capabilities of \ac{bs}'s signals, \ac{isac}-based \ac{mdm} is promising to enhance the observation region and improve monitoring flexibility.
% However, due to the duplex nature of \ac{isac}, \ac{mdm} incurs significant challenges in sensing signal extraction/separation, yielding traditional \ac{mdd} estimation techniques, which are designed for accelerometer/radar-based counterparts, inefficient or even inapplicable.
However, due to the poor sensing quality of \ac{isac}, \ac{mdm} incurs significant challenges in sensing signal extraction/separation.
This limitation makes traditional \ac{mdd} estimation techniques, which are designed for accelerometer/radar-based counterparts, inefficient or even inapplicable.

The primary goals of this study are to: 1) introduce an \ac{isac}-based \ac{mdm} architecture, 2) propose a learnable template matching approach for \ac{mdm}, and 3) demonstrate the \ac{mdd} estimation performance by simulations and experiments.

\subsection{Related Work}

Many efforts have been dedicated to enhancing \ac{isac} systems for sensing functionality, including optimizing sensing performance through transmitter/receiver design \cite{huang2020majorcom,ma2021spatial,liu2024hybrid,hua2023optimal,tsinos2021joint} and studying the limits of sensing capabilities \cite{liu2021cramer,sadeghi2021target,wang2019first,feng2020joint,ma2020joint,liu2022integrated}.
Transmitter/receiver design relies on resource allocation optimization strategies, thereby striking a better trade-off between communication and sensing on a shared platform.
A commonly adopted strategy in transmitter/receiver configuration is to design temporally/spectrally/spatially orthogonal waveforms \cite{huang2020majorcom,ma2021spatial,tsinos2021joint} or optimize their beamformer \cite{liu2024hybrid,hua2023optimal} to maximize the sensing quality while ensuring fundamental communication. 
However, these techniques fail to achieve sensing tasks, as the sensing capabilities in \ac{isac} systems are still inadequate compared to those of dedicated radar systems. 
The authors of \cite{liu2021cramer} employed the Cramér-Rao bound to analyze the sensing limits of sensing in \ac{isac} systems. 
Furthermore, in \cite{feng2020joint,ma2020joint,liu2022integrated}, the authors offered insights into the sensing aspects of \ac{isac} systems, highlighting the challenges and vital technologies involved in enhancing the sensing capabilities of these systems.
{\color{black}The existing approaches to addressing the \ac{isac} problems primarily remain on improving/enhancing the communication performance or sensing evaluation metric, such as \ac{snr} and \ac{sinr} \cite{liu2024hybrid,liu2021cramer}.
These designs limit the spectrum and aperture flexibility of sensing signals compared to standard radar systems, failing to fully resolve the core issues in sensing operations \cite{sadeghi2021target,wang2019first}. }

With the development of model-based deep learning, \ac{ai}-assisted approaches have attracted significant attention from researchers for addressing challenges in the signal/image processing domains \cite{monga2021algorithm,liu2023interpretable,fuhg2022physics,hughes2019wave,li2023x}.
{\color{black}The superiority of \ac{ai}-assisted approaches has been demonstrated in various systems related to sensing tasks \cite{chen2020review,guo2021physics,li2018deepnis,deshmukh2022physics,solomon2019deep,eldar2009robust,li2020efficient}. 
Recently, model-based deep learning frameworks have been widely explored for reconstruction problems, where physical models or structural priors are embedded into neural networks to improve interpretability and performance \cite{li2023model,li2024model,li2025enhanced}.}
For instance, the model-based methods were reviewed in \cite{monga2021algorithm} for solving standard signal processing challenges, and authors in \cite{liu2023interpretable} especially introduced the principles and applications of model-based deep learning in \acf{em} signal processing. 
To accurately model the \ac{em} characteristics of sensing signals, the authors in \cite{chen2020review,guo2021physics,li2018deepnis} outlined the fundamental approach to establishing the inverse scattering problem model and suggested a high-precise inverse scattering solution based on physics-assisted \ac{ai}. 
Considering the cluttered environment, achieving signal deblurring/enhancing and clutter suppression is necessary. 
To optimize signal quality, \cite{li2020efficient} proposed a deep blind image deblurring technique via algorithm unrolling, directly improving the sensing quality by \ac{ai}-assisted approaches. 
Furthermore, a low-rank-based clutter model was proposed in \cite{solomon2019deep,eldar2009robust} to indicate the design of the network, enabling effective clutter suppression in image domains. 
These \ac{ai}-assisted approaches demonstrate superior performance by combining the advantages of deep learning and physics constraints \cite{monga2021algorithm,liu2023interpretable}.
This combination provides a promising solution to the low sensing quality challenges posed by limited spectrum and aperture in \ac{isac} systems.

\subsection{Motivation and Contribution}

Most existing studies on \ac{isac} systems have focused on communication tasks \cite{huang2020majorcom,ma2021spatial,liu2024hybrid,hua2023optimal,tsinos2021joint,liu2021cramer,sadeghi2021target,wang2019first,feng2020joint,ma2020joint,liu2022integrated}, whereas weak target sensing, particularly the \ac{mdm} problem, has not been explored yet. 
Although the sensing problem, such as synthetic aperture radar (SAR) imaging, has been studied in \cite{wang2019first}, the sensing signals remain restricted by the communication data. 
Therefore, their proposed schemes are inapplicable and sufficient to achieve the \ac{mdm} under limited sensing quality.
{\color{black}Due to these challenges, exploring the \ac{isac}-based \ac{mdm} approach, which aims to achieve satisfactory performance under poor sensing quality, remains in its infancy.}

To fill this gap, we focus on the \ac{isac}-based \ac{mdm} system, wherein the \ac{bs} performs both communication and sensing tasks. To be specific, the \ac{bs} simultaneously sends the symbols to communicate with users and transmits sensing signals to monitor infrastructure. However, due to limitations in the shared \ac{bs} platforms, the \ac{mdd} estimation from the proposed \ac{mdm} system encounters significant challenges, primarily due to low resolution and clutter interference. To address these issues, we begin by modeling the environment clutter as a combination of deterministic vibration signals and stochastic noise, representing the \ac{bs}'s coverage region. The \ac{isac}-based \ac{mdd} extraction problem is then formulated as a reconstruction problem. Thanks to the engagement of deep learning, we propose a learnable template matching approach for solving this problem. First, the \ac{cnn}-based network is developed to overcome phase wrap challenges in \ac{em} signal processing. In addition, we propose an \ac{ai}-based clutter suppression model that reforms clutter suppression as a target signal enhancement task. We then integrate a learnable template block within the neural network layers to separate \ac{mdd} signal features and guide the network training process.
Numerical and experimental results demonstrate that our proposed \ac{ltm} achieves impressive accuracy in \ac{mdm} estimation.

The main contributions of this work are summarized as:
\begin{itemize}
    \item \textbf{\Ac{isac}-based \ac{mdm} System}: 
    To the best of our knowledge, we are the first to propose an \ac{isac}-based \ac{mdm} system, where the \ac{bs} enables communication with users and transmission and reception of the sensing signals. 
    Specifically, we study the sensing signal extraction and separation techniques, achieving \ac{mdd} estimation in the proposed system.
    
    \item \textbf{Clutter-Target Feature Model}: 
    We convert the clutter suppression problem into a sensing signal enhancement task, facilitating the efficient mitigation of complex environmental clutter.
    Leveraging the periodic motion characteristics of the \ac{mdd} signals, the \ac{mdd} estimation is formulated as a reconstruction problem involving coupled deterministic periodic motion and unknown clutter.

    \item \textbf{LTM Network Design}:
    We propose a \acf{ltm} approach for solving the reconstruction problem by splitting it into phase unwrapping and signal decoupling tasks. 
    To manage the wrapped phase in the sensing signal, we employ a \ac{cnn}-based method specifically designed to handle periodic phase signals. 
    To address the issue of coupled \ac{mdd} and clutter signals, we propose a novel architecture that integrates a learnable periodic motion template to enhance the \ac{mdd} and the hidden layers of \ac{cnn} as a clutter filter.

    \item \textbf{Performance}:
    Numerical and experimental results demonstrate superior performance in \ac{mdd} estimation and the effectiveness of our proposed approach.
    % Specifically, the \ac{mdm} of the real \ac{bs} measurement achieves $99$\% $2$~mm pressure detection accuracy within $0.2$~mm error and $1$~mm pressure detection accuracy within $0.2$~mm error.
    Furthermore, the simulation-trained \ac{ltm} method exhibits the function of separating \ac{bs} vibrations from infrastructure vibrations, confirming the signal decoupling capability of our proposed approach.
    
\end{itemize}

\subsection{ Organization and Notation}

The rest of this paper is organized as follows: Section \ref{system model} presents the \ac{isac}-based \ac{mdm} system, reviews the operational mode of the \ac{isac}, and formulates both the \ac{mdd} forward model and the \ac{mdd} estimation problem within our \ac{isac} architecture. Section \ref{solution} presents efficient methods to extract sensing signals from the \ac{bs} and separate the \ac{mdd} signal from these sensing signals. Section \ref{sec:Sims} numerically demonstrates the solution of this proposed system, providing performance evaluations through simulations and experiments in \ac{isac} settings. Finally, Section \ref{sec:Conclusions} concludes the paper.

Throughout the paper, we use boldface lower-case and upper-case letters for vectors and matrices, respectively. The $(x,y)$-th element of the matrix $\boldsymbol{A}$ is denoted by $\boldsymbol{A}(x,y)$. The $\ell_2$ norm, $\ell_1$ norm, conjugate operation, transpose, stochastic expectation, convolution product, and Fourier transform are written as $\Vert\cdot \Vert_2$, $\Vert\cdot\Vert_1$, $(\cdot)^*$, $(\cdot)^T$, $\mathbf{E}(\cdot)$, $*$, and $\mathscr{F}( \cdot )$, respectively. We use $\boldsymbol{I}_N$ to denote an $N$-dimensional
identity matrix, $\boldsymbol{0}_{M\times N}$ is an $M\times N$ zero matrix, and $\mathbbm{C}$ is the complex set.

\section{System Model}
\label{system model}

\subsection{Sensing principles of ISAC systems}

\begin{figure}[htbp]
    \centering
    \includegraphics[width=.95\linewidth]{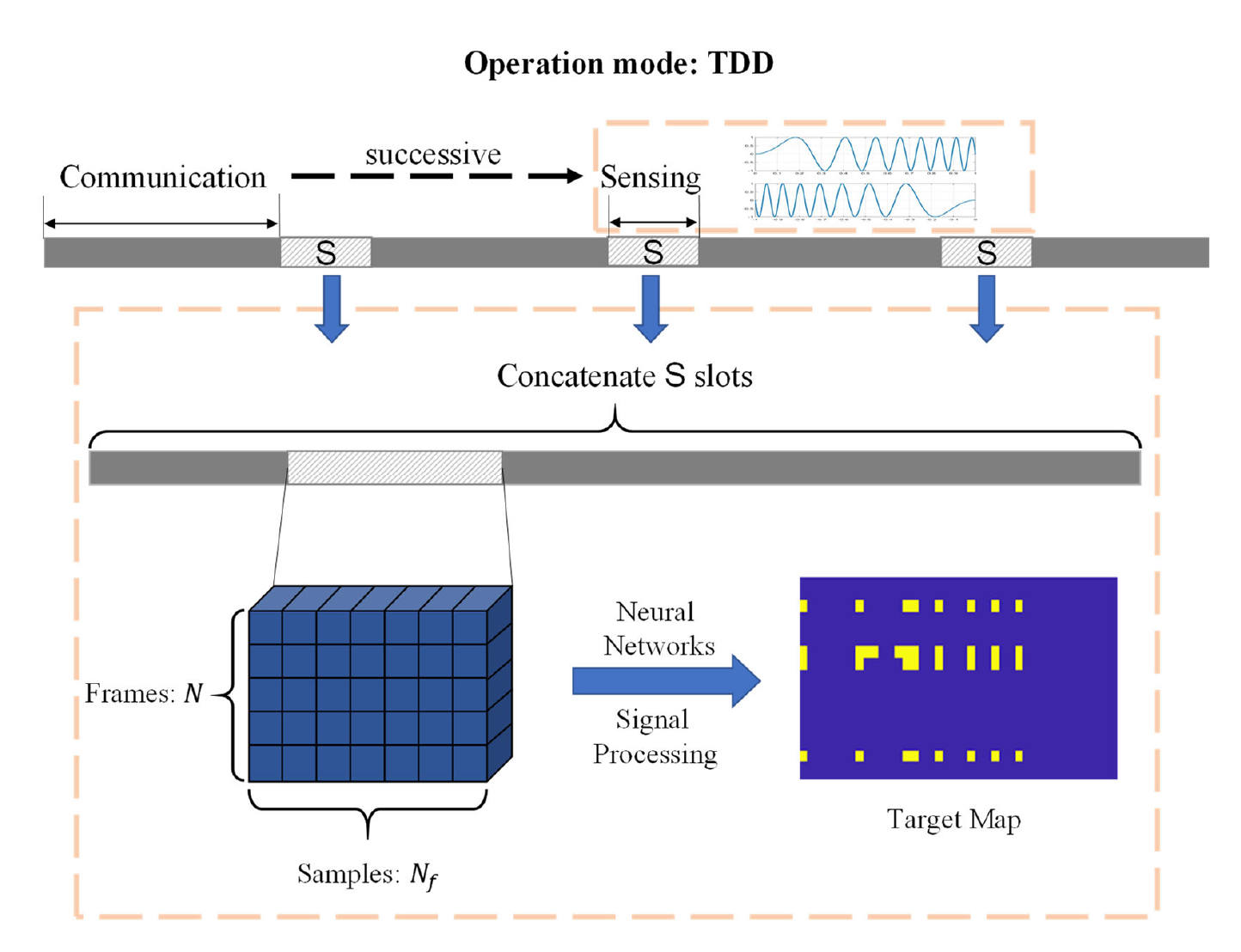}
    \caption{The entire operation of the ISAC system, involves strategies for both the transmit communication modulation and the sensing signal modulation.}
    \label{fig:tdd}
\end{figure}

We consider a \ac{bs}-based \ac{isac} system, where the deployed \ac{bs} consists of a pair of adjacent transmit and receive antennas. 
In this setup, both the transmit and receive units are connected to a unified data control center via cables and are centrally managed by the control center. 
The system operates in \ac{tdd} mode to facilitate \ac{isac}, as illustrated in Fig.~\ref{fig:tdd}. 
Specifically, one transmit antenna alternates between performing radar sensing and communication functions, while another antenna is dedicated to continuously receiving radar echoes. 

In Fig.~\ref{fig:tdd}, the communication symbols are generated using 16-quadrature amplitude modulation (QAM), while the sensing waveforms are modulated as linear frequency modulated continuous wave (FMCW) signals. 
This work focuses primarily on achieving high-quality sensing within the framework of \ac{isac} systems. 
Without loss of generality, we assume that communication performance is fully guaranteed and the communication symbols are uncorrelated with the sensing signals.
In addition, we concatenate the time slots $\mathcal{S}$ allocated for sensing into a single observation, where each time slot contains $N$ frames with a sampling rate of $N_f$ per frame.

Let $B$ denote the bandwidth of the sensing signals and $f_0$ be the starting frequency. Based on the principles of the FMCW signals, {\color{black} the \ac{bs}'s received echoes from the $n$-th sample of the $l$-th frame}, for $l=1,...,N$, is given by
\begin{equation}
    \begin{aligned}
        \centering
        \boldsymbol{S}(l,n) = \exp{\left(-j2\pi \left(f_0+\frac{B}{N_f}n\right)\frac{\boldsymbol{r}(l)}{c_0}\right)},~n=1,...,N_f,
    \end{aligned}
    \label{fmcw}
\end{equation}
where $c_0$ is the speed of light, and $\boldsymbol{\tau}(l)=\boldsymbol{r}(l)/{c_0}$ is the time delay of the echo propagating between the transceiver and the target with $\boldsymbol{r}(l)$ being the round-trip distance.
Equation (\ref{fmcw}) is mathematically consistent with that of the standard FMCW system, allowing us to operate the \ac{isac} platform for target sensing by emulating the FMCW processing architecture.

\subsection{ISAC-based mDM Architecture}
\label{sec:isacbvm}

\begin{figure}[htbp]
    \centering
    \includegraphics[width=.95\linewidth]{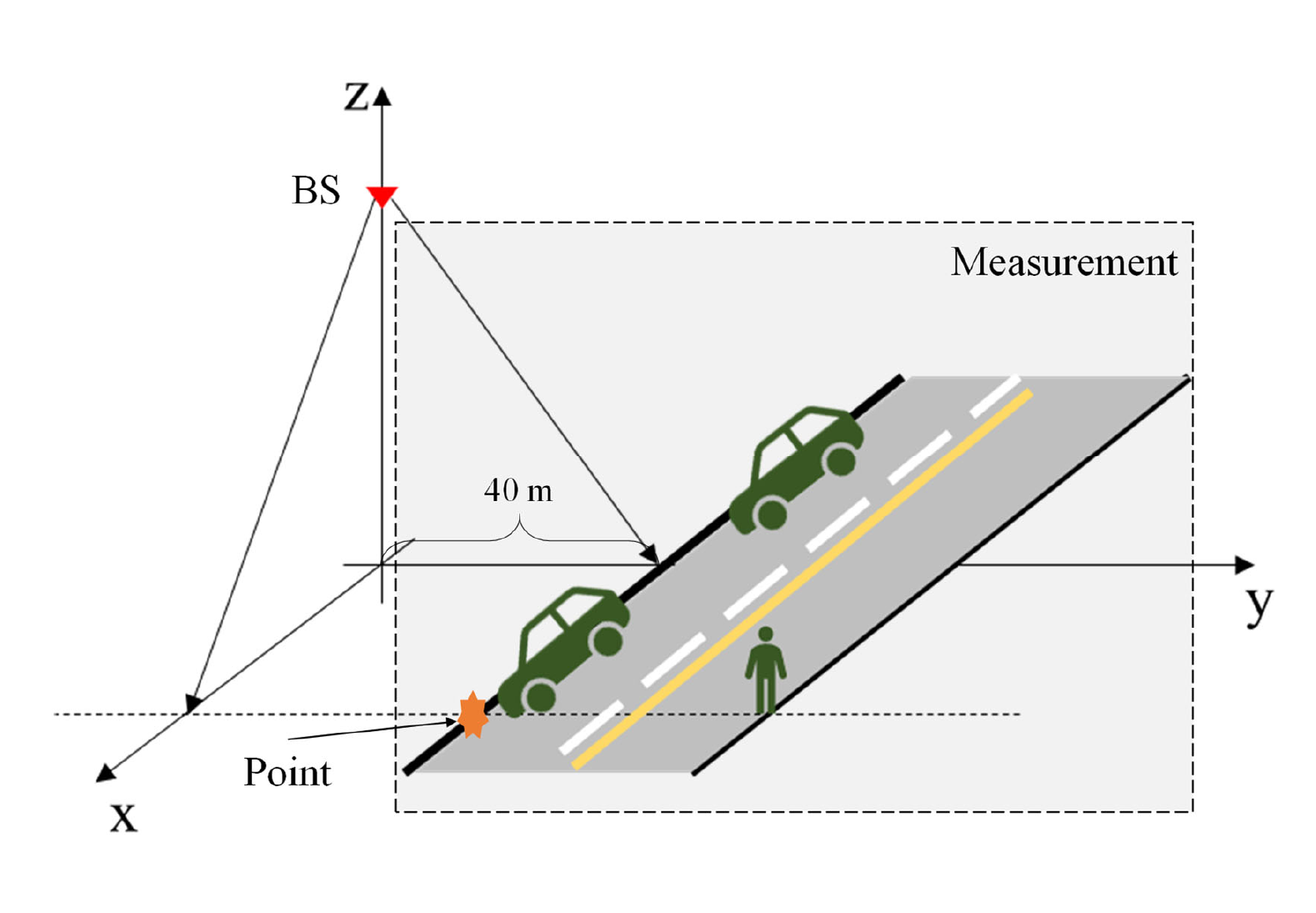}
    \caption{The geometry of the ISAC-based mDM system.}
    \label{fig:bvm}
\end{figure}

As shown in Fig.~\ref{fig:bvm}, we consider target detection under the constrained sensing capability in the \ac{isac} system, with a particular focus on the \ac{mdm} task. 
The \ac{isac}-based \ac{mdm} system aims to estimate the \ac{mdd} within a dynamic environment. 
Specifically, the \ac{bs} is responsible for illuminating the infrastructure scenario and receiving echoes scattering from vehicles and pedestrians on the road. 
According to (\ref{fmcw}), the signals received by the \ac{bs} are modulated by targets located at different distances. 
Let the observation target be treated as the point scatterer and vehicles and pedestrians be clutter. 
We suggest to characterize received echoes from the urban scenario as a combination of target signals, clutter signals, and background noises. 
The combined received signal is expressed as
\begin{equation}
    \begin{aligned}
        \centering
        \boldsymbol{S}(l,n) =& \underbrace{\exp{\left(-j2\pi \left(f_0+\frac{B}{N_f}n\right)\frac{\boldsymbol{r}(l)}{c_0}\right)}}_{\rm Target~Signal}\\
        &+\underbrace{\sum_{i=1}^{Q}\exp{\left(-j2\pi \left(f_0+\frac{B}{N_f}n\right)\frac{\boldsymbol{\mathcal{H}}(i,l)}{c_0}\right)}}_{\rm Clutter~Signal}+\boldsymbol{V}(l,n),\\
        &~~~~~~~~~~~~~~~~~~~~~~~~~~~~~~~~~~~~~n=1,...,N_f,
    \end{aligned}
    \label{sense}
\end{equation}
where $\boldsymbol{V}$ is the complex \ac{awgn}, $\boldsymbol{r}(l)$ and $\boldsymbol{\mathcal{H}}(i,l)$ denote the round-trip distances at the $l$-th frame for the target and the $i$-th clutter element in the clutter group, respectively.
Note that the distance of interest at the observation point $\mathrm{P}$ on the infrastructure is affected by the vibration of the \ac{bs} and the coupled clutter signals.
Therefore, the round-trip distance from $\mathrm{P}$ to the \ac{bs} extracted from (\ref{sense}) can be represented as 
\begin{equation}
    \begin{aligned}
        \centering
        % \boldsymbol{d} = \boldsymbol{r}(1) + \underbrace{d_b + d_a \sin \left(2\pi f_v \frac{l-1}{N}\right)}_{\boldsymbol{d}_v} + \boldsymbol{h}_v ,~l=1,...,N,
        \boldsymbol{d}(l) \triangleq \boldsymbol{r}(1) + \boldsymbol{d}_v(l) + \boldsymbol{h}_v(l) ,~l=1,...,N,
    \end{aligned}
    \label{distance}
\end{equation}
where the $\boldsymbol{r}(1)$ is the initial distance for $\mathrm{P}$, $\boldsymbol{d}_v$ is the infrastructure deformation caused by loading vehicles, and $\boldsymbol{h}_v$ accounts for the distance distortion due to clutter and noise.
% The main purpose of \ac{isac}-based \ac{mdm} system is to achieve \ac{mdd} estimation from the sensing signals provided by the \ac{isac} systems. 

To support the precise \ac{mdd} estimation, the distance corresponding to $\mathrm{P}$ of the sensing signal in (\ref{distance}) needs to be extracted and separated from the clutter interference and background noise.
Similar to \cite{tian2019vibration}, we first model the \ac{mdm} signal as the standard periodic wave
\begin{equation}
    \begin{aligned}
        \centering
        \boldsymbol{d}_v =d_r + d_a \sin \left(2\pi f_v \frac{l-1}{N}\right),~l=1,...,N,
    \end{aligned}
    \label{bvdsig}
\end{equation}
{\color{black}where $d_r$ is the radial distance between $\mathrm{P}$ and the \ac{bs}, as known as \ac{mdd}. $d_a$ and $f_v$ represent the vibration amplitude and the vibration frequency of $\mathrm{P}$, respectively.}

As previously discussed, the environment clutter consists of the \ac{bs} vibration and distance distortion induced by loading vehicles.
Thus, the environment clutter can be characterized as a combination of deterministic \ac{bs}'s vibrations and the stochastic noise. 
For simplicity, we assume that the vibration of \ac{bs} follows a stable periodic wave pattern, characterized by an amplitude $d_b$ and a frequency $f_b$. 
The distance distortion caused by environmental clutter signals can be expressed as 
\begin{equation}
    \begin{aligned}
        \centering
        \boldsymbol{h}_v = d_b \sin \left(2\pi f_{b} \frac{l-1}{N}\right) + \boldsymbol{v}, ~l=1,...,N,
    \end{aligned}
    \label{clutter}
\end{equation}
{\color{black} where $\boldsymbol{v}$ is the \ac{awgn}.}

Inspired by other \ac{mdm} works \cite{yang2023multi,sadeghi2023using,ma2023structural}, the initial distance for $\mathrm{P}$ can be eliminated through direct path compensation.
Upon removing the direct path, and combining (\ref{bvdsig}) with (\ref{clutter}), (\ref{distance}) is then represented as 
% $\boldsymbol{d} = \boldsymbol{d}_v+\boldsymbol{h}_v$.
\begin{equation}
    \begin{aligned}
        \centering
        \boldsymbol{d} = d_r &+ d_a \sin \left(2\pi f_v \frac{l-1}{N}\right) \\
        &+d_b \sin \left(2\pi f_{b} \frac{l-1}{N}\right) + \boldsymbol{v},~l=1,...,N.&
    \end{aligned}
    \label{bvmsig}
\end{equation}

\subsection{Problem formulation}

In considered \ac{isac}-based \ac{mdm} system, we aim to extract the sensing signal in (\ref{distance}) from the received echoes, achieving high-precise \ac{mdd} estimation. 
In particular, we extract the round-trip distance from (\ref{sense}) by detecting the phase of the observation point, expressed as $\boldsymbol{d}(l)=\angle\mathscr{F}\left(\boldsymbol{S}(l,n)\right)$.
From (\ref{bvdsig}) and (\ref{clutter}), the radial distance $d_r$, vibration amplitude $d_a$, and frequency $f_v$ are \ac{mdd} parameters, while the entire term $\boldsymbol{h}_v$ represents clutter that we want to remove.
As the \ac{mdm} signal is aliased with clutter signals, the \ac{mdd} estimation requires decoupling the \ac{mdd} signal from the sensing signals.
To address this challenge, we employ optimization approaches to reconstruct the \ac{mdm} signal in (\ref{bvdsig}) from the sensing signal in (\ref{bvmsig}).
Specifically, we propose to find feasible solutions that minimize the loss of \ac{mdd} reconstruction while ensuring effective clutter suppression.
The resulting problem w.r.t. the \ac{mdd} reconstruction is formulated as
\begin{subequations}
    \begin{align}
        \centering
        \mathop{\arg\min}_{d_r, d_a, f_v} &\ \frac{1}{2} \Vert \boldsymbol{d} - \boldsymbol{d}_v - \boldsymbol{h}_v \Vert_2^2 + \lambda \mathcal{R}\left(\boldsymbol{h}_v\right),\\
        {\rm s.t.} &\ \boldsymbol{d}_v =d_r + d_a \sin \left(2\pi f_v \frac{l-1}{N}\right),~l=1,...,N,\label{dv}\\
        &\ \boldsymbol{h}_v = d_b \sin \left(2\pi f_{b} \frac{l-1}{N}\right) + \boldsymbol{v}, ~l=1,...,N \label{hv},
    \end{align}
    \label{P0}
\end{subequations}
where $\lambda$ is the penalty part for the balance of the clutter suppression and fidelity to the \ac{mdd} reconstruction.
(\ref{P0}) presents the regression problem with the primary goal of estimating the parameters of the \ac{mdd} with periodic characteristics as defined in (\ref{dv}). 
Specifically, we characterize the \ac{mdd} signal as a single-frequency sinusoidal signal with a fixed displacement bias. 
The objective of (\ref{P0}) is to estimate the bias and frequency parameters of a periodic type of signal. 
In addressing clutter interference, $\mathcal{R}\left(\boldsymbol{h}_v\right)$ denotes the clutter suppression filter for the \ac{mdm} task, which needs to be designed according to the environmental clutter model described in (\ref{hv}).

Due to the phase-wrapping nature of the received signal at the \ac{bs}, the extracted sensing signal $\boldsymbol{d}$ via angle estimation is nonlinear. 
Moreover, \ac{isac}-based \ac{mdm} systems constrained by low sensing quality operate in complex conditions, resulting in challenges in addressing this problem.

\section{mDD Estimation Solution in ISAC-based mDM System}
\label{solution}

To overcome these difficulties, we reformulate the clutter suppression problem as the \ac{mdd} signal enhancement task and propose a \ac{ltm} approach to address the \ac{mdd} estimation issue.
Specifically, the \ac{cnn}-based method is suggested to solve the phase wrap obstacles, while the \ac{ltm} network architecture is designed to decouple the clutter and \ac{mdd} signals. 

We begin with introducing the clutter-target feature model in Section~\ref{sec:clu-tar}, which we then utilize to design the \ac{ltm} network architecture in Section~\ref{sec:ltm}.
Finally, we summarize the \ac{ltm} approach and loss function design in Section.~\ref{sec:loss}. 

\subsection{Clutter-Target Feature Model}
\label{sec:clu-tar}

Given the dynamic nature of environmental clutter signals, obtaining a high-fidelity clutter model for the \ac{isac}-based \ac{mdm} scenario poses significant challenges. 
Nevertheless, the target signal feature of interest adheres to a standard model, which can be mathematically expressed as (\ref{bvdsig}). 
Inspired by this characteristic, we aim to enhance the target feature while overlooking clutter interference, focusing solely on extracting the relevant target information. 
% To address this, template matching techniques are employed to separate the \ac{mdm} signal from the coupled sensing signal, as detailed below. 
% We begin with the template model and reformulate the original problem (\ref{P0}) in terms of the clutter-target feature transformation.

{\color{black}Let $\boldsymbol{\hat{d}}_v$ denote the learnable \ac{mdm} signal, and let $\hat{d}_r$ and $\hat{d}_a \triangleq \alpha\hat{d}_r$ represent the \ac{mdd} signal and vibration amplitude, respectively.
% Without loss of generality, the vibration amplitude $\hat{d}_a$ is determined by the structure deformation $\hat{d}_r$ \cite{pramudita2023fmcw,zhao2022dynamic}.
With the correlation coefficient $\alpha$ empirically set to $0.1$, the target feature template with the length of $N_s$ can be expressed as 
\begin{equation}
    \begin{aligned}
        \centering
        \boldsymbol{\hat{d}}_v = \hat{d}_r + \alpha \hat{d}_r \sin \left(2\pi \hat{f}_v t_i\right),~i=1,...,N_s,
    \end{aligned}
    \label{template}
\end{equation}
where the $\hat{f}_v$ is the frequency of the template signal, and the sample rate during one slot $t_i = (i-1)/N_s$.

% The primary concept of template matching is to determine the peak value and the number of peaks to estimate the displacement and frequency, respectively  \cite{zhou2021robust}. 
In this case, the correlated signal, such as the \ac{mdd} signal, will be enhanced, while the non-standard signal, such as clutter, will be weakened. 
% In the \ac{isac}-based \ac{mdm} scenario considered here, the sensing signal, expressed as (\ref{bvmsig}), is suggested to be matched with the template in (\ref{template}). 
In particular, the \ac{mdd} parameters are obtained through the convolution process of the sensing signal with the template.
% This process effectively mitigates the clutter interference by leveraging the inherent correlation properties of convolution. 
Thus, the resulting problem can be reformulated as:
\begin{equation}
    \begin{aligned}
        \centering
        \mathop{\arg\max}_{d_r, f_v} &\ \underbrace{\Vert  \boldsymbol{d}*\boldsymbol{\hat{d}}_v  \Vert_2^2}_{\rm Target~Feature} - \lambda_1 \underbrace{\mathbf{E}\left(\Vert \boldsymbol{d} - \boldsymbol{\hat{d}}_v \Vert_2^2\right)}_{\rm Clutter ~Suppression},
    \end{aligned}
    \label{P1}
\end{equation}
where $\lambda_1$ is the penalty parameter for balancing the target and clutter signals.
As illustrated in (\ref{P1}), the correlated convolution and stochastic expectation operators are employed for target signal enhancement and clutter suppression, respectively.

However, template matching is conducted in the round-trip distance domain, where phase-wrapping issues still exist.
In addition, the fixed template in (\ref{template}) overlooks the spectral flexibility of the \ac{mdd}, leading to inadequate \ac{mdd} separation and estimation performance.
These limitations motivate the proposed \ac{ltm} framework, which integrates phase preprocessing and adaptive template learning.}

\subsection{LTM Network Design}
\label{sec:ltm}

\begin{figure*}[htbp]
    \centering
    \includegraphics[width=.95\linewidth]{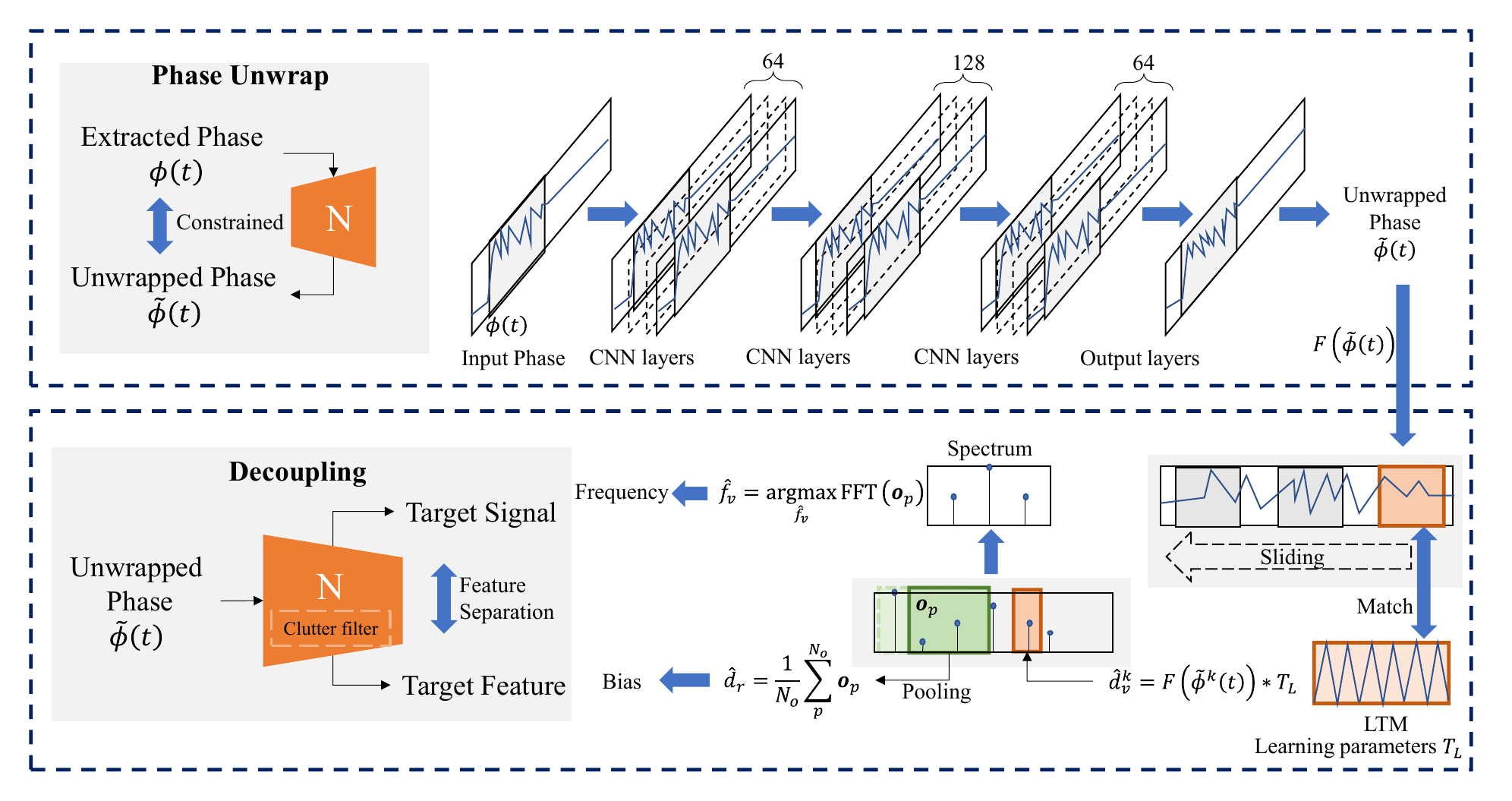}
    \caption{LTM Architecture: The architecture consists of two main components. The top block represents the workflow for the CNN-based phase unwrapping, which addresses the phase-wrapping issues in the extracted signals. The bottom block illustrates the LTM network, specifically designed for decoupling the mDD signal from clutter interference.}
    \label{fig:ltm}
\end{figure*}

We propose an \ac{ltm} network to address the \ac{mdd} estimation task by splitting the original problem into two tasks: phase unwrapping and signal decoupling, as shown in Fig.~\ref{fig:ltm}. 
This approach leverages the superiority of \ac{cnn} and learnable templates to effectively manage wrapped phases and decouple signals in our considered \ac{mdm} scenario.
In the following, we provide a detailed introduction to the \ac{cnn}-based phase unwrapping and \ac{ltm} neural network for the signal decoupling in detail.

\subsubsection{Phase Unwrapping via CNN-Based Method}
The phase-wrapping effect is the essential challenge in periodic \ac{em} signal processing, which appears due to the modulo nature of the observed phase.
To address this issue, we employ the \ac{cnn}-based method tailored to unwrap the phase of \ac{bs}'s received signals.

{\color{black}As illustrated in the top block of Fig.~\ref{fig:ltm}, we first extract the phase from the \ac{bs}'s received signal and design the network layers for phase unwrapping.
The proposed \ac{cnn}-based architecture consists of multiple convolutional layers designed to extract spatial features from the \ac{mdd} data sequentially.
The unwrapping operator is expressed as
\begin{equation}
    \begin{aligned}
        \boldsymbol{\widetilde{\phi}}(t) = \mathcal{F}_N \left(\mathcal{W},\boldsymbol{\phi}(t)\right),~\forall t \in \frac{l-1}{N}, 
    \end{aligned}
\end{equation}
where the $\mathcal{F}_N (\cdot)$ is the multi-layer \ac{cnn}, $\boldsymbol{\phi}(t)$ is the phase term of the sensing signal extracted from the \ac{bs}, and $\boldsymbol{\widetilde{\phi}}(t)$ is consequently the corresponding unwrapped phase signal after \ac{cnn} parts. Without loss of generality, the phase term extracted from the \ac{bs} is assumed to exhibit continuity. Thus, the multi-layer operator $\mathcal{F}_N (\cdot)$, which maps the raw phase to the unwrapped phase, is subject to the consistency constraint. 
% This inherent physical limitation informs the design of the activation function in the \ac{cnn}-based phase unwrapping module. 
To be concrete, the nonlinear activation function in $\mathcal{F}_N (\cdot)$ is designed to ensure both the continuity of the phase unwrapping and the consistency of the phase-deformation mapping. As shown in Fig.~\ref{fig:ltm}, the proposed \ac{ltm} approach performs the multi-layer \ac{cnn} to extract the phase signal to the unwrapped phase term. Specifically, the proposed network processes the one-dimensional time-domain phase signal through multiple convolutional layers, transforming the raw input into a high-dimensional representation. Then, we employ the fixed-length convolutional kernels across all \ac{cnn} layers to extract and expand the phase features, ensuring effective preprocessing for phase unwrapping. }

\subsubsection{Signal Decoupling via Learnable Template}

Signal decoupling is a key issue in separating distinct signal components in \ac{isac} \ac{mdm} systems, especially when processing complex received echoes. 
To achieve this, we propose constructing a learnable template, which is adaptively acquired while training to capture the \ac{md} signatures of both the target and clutter. 
The learnable template enables effective decoupling of the target signal from the environment by leveraging its unique spectral and spatial characteristics. 
Specifically, the network incorporates time-limited templates and complex Fourier transform operations to extract the bias and frequency components of the target signal. 
The proposed signal decoupling comprises two main components: target signal enhancement and clutter suppression, as illustrated in Fig.~\ref{fig:ltm}.
We begin by introducing the \ac{ltm}-based target signal enhancement.
{\color{black}By replacing the fixed template $\boldsymbol{\hat{d}}_v$ in (\ref{P1}) with the learnable parameters $\boldsymbol{T}_L$, the time-spatial matching process can be represented as}
\begin{equation}
    \begin{aligned}
        \centering
        \boldsymbol{\hat{d}}^k_v = \mathcal{F}_N \left(\boldsymbol{\widetilde{\phi}}^k\right)*\boldsymbol{T}_L,
    \end{aligned}
    \label{ts-match}
\end{equation}
where the length of $\boldsymbol{T}_L$ is defined as $N_s$, and the operator $\mathcal{F}_N(\cdot)$ is utilized for mapping the phase into deformation signal. Moreover, the deformation displacement $\boldsymbol{d}^k=\mathcal{F}_N \left(\boldsymbol{\widetilde{\phi}}^k\right)$ is defined as deformation displacement corresponding unwrapped phase $\boldsymbol{\widetilde{\phi}}^k$, which remains the same signal length as the template $\boldsymbol{T}_L$.
Following the \ac{ltm} process in (\ref{ts-match}), the estimated displacement output $\boldsymbol{o}$ is processed through a sliding-window operation with the learnable parameters $\boldsymbol{T}_L$, expressed as
\begin{equation}
    \begin{aligned}
        \centering
        \boldsymbol{o} = \left[\boldsymbol{\hat{d}}^1_v,...,\boldsymbol{\hat{d}}^{N_L}_v\right],
    \end{aligned}
    \label{o}
\end{equation}
where the sliding step is denoted as $\Delta_x$, and the total number of signal components is $N_L$. Specifically, to ensure the continuity of the displacement signal, the length of the learnable template and the sliding step satisfy the following constraints: $\Delta_s \ll N_s$.

Subsequently, we describe the pipeline for decoupling the target signal from the clutter signal. The key concept of signal decoupling involves extracting matched signals with the template $\boldsymbol{T}_L$ from each entry within $\boldsymbol{o}$, while effectively suppressing unmatched clutter signals. Inspired by the max pooling properties in the PyTorch architecture, we utilize the max pooling process to extract interested feature points. In particular, we apply the max pooling operator to the estimated displacement in (\ref{o}) to extract the peak values from each signal segment $\boldsymbol{\hat{d}}^k_v$. Let the max pooling operator be $\mathcal{P}_M(\cdot)$, the \ac{ltm}-based feature extraction can be expressed as
\begin{equation}
    \begin{aligned}
        \centering
        \boldsymbol{o}_{\mathcal{P}} = \mathcal{P}_M\left(\boldsymbol{\hat{d}}^k_v\right)=\max_i{\boldsymbol{o}^i},~i=1,...,N_o,
    \end{aligned}
    \label{maxpool}
\end{equation}
where the $\boldsymbol{o}^i$ is the displacement segment of length $N_p$, and $N_o$ is the length of feature signal $\boldsymbol{o}_{\mathcal{P}}$. 
Obviously, the feature signal output reaches a peak if and only if the input displacement segment $\boldsymbol{\hat{d}}^k_v$ aligns precisely with the template $\boldsymbol{T}_L$, which can be captured by the max pooling operator $\mathcal{P}_M(\cdot)$.
In addition, the peaks captured by the pooling layer directly correspond to the periodic displacement components of the \ac{mdd} signal, while the unmatched clutter signals are concurrently eliminated. To this end, (\ref{maxpool}) achieves simultaneous extraction of the \ac{mdd} signal and suppression of clutter.
However, due to the scaling effects of the pooling layer, there exists an estimation error between the extracted peaks and the real displacement of the \ac{mdd} signal. This estimation error primarily comes from the inappropriate design of the pooling layer size $N_p$ and the pooling stride $\Delta_p$. To adequately capture the \ac{mdd} signal characteristics, the pooling stride ensures the Nyquist-like sampling conditions for the output signal $\boldsymbol{o}$, satisfying: $\Delta_p \leq N_s/2$.
Furthermore, to constrain the output signal's frequency and effectively suppress high-frequency clutter, the pooling kernel length $N_p$ adheres to: $N_p \geq 2N_s$.
Finally, the estimated \ac{mdd} is calculated by,
\begin{equation}
    \begin{aligned}
        \centering
        \hat{d}_r = \frac{1}{N_o}\sum_{i=1}^{N_o}\mathcal{P}_M^i \left(\boldsymbol{\hat{d}}^k_v\right).
    \end{aligned}
    \label{dr_esti}
\end{equation}

As previously described in Section.~\ref{sec:clu-tar}, the template follows the harmonic vibration model, and the estimated \ac{mdm} signal $\boldsymbol{\hat{d}}^k_v$ subsequently exhibits single-frequency vibration characteristics within the observation time window. Leveraging these inherent frequency characteristics of the \ac{mdm} signal, we propose embedding the \ac{fft} into the neural network, thereby establishing a unified time-frequency-spatial framework for \ac{mdd} estimation. 
In the time-spatial domain, we design a novel max pooling layer to extract the \ac{mdd} signal via (\ref{dr_esti}). Concurrently, from the perspective of the frequency analysis, we embed the \ac{fft} module to capture the vibration frequency component of the \ac{mdm} signal, expressed as 
\begin{equation}
    \begin{aligned}
        \centering
        \mathcal{F}\left(\boldsymbol{\hat{d}}^k_v \right) = \arg \max _{\hat{f}_v} \left|\mathscr{F}\left(\boldsymbol{\hat{d}}^k_v\right)\right|.
    \end{aligned}
    \label{fre_esti}
\end{equation}
However, due to the non-differentiable nature of the frequency index selection, directly embedding the operator $\mathscr{F}(\cdot)$ poses implementation challenges within the backpropagation framework of neural networks. To overcome this limitation, we propose a spectrum power-based optimization strategy that leverages the intensity differences of frequency components in the spectrogram between the target and clutter signals. Under this strategy, the target signal is identified and extracted based on its distinct frequency components, whereas clutter signals, characterized by relatively flat spectra, are effectively suppressed.
This approach enhances the ability to distinguish the target from environmental noise, improving overall system performance.

\subsection{Network Architecture and Loss Function Design}
\label{sec:loss}
\begin{figure}[htbp]
    \centering
    \includegraphics[width=0.95\linewidth]{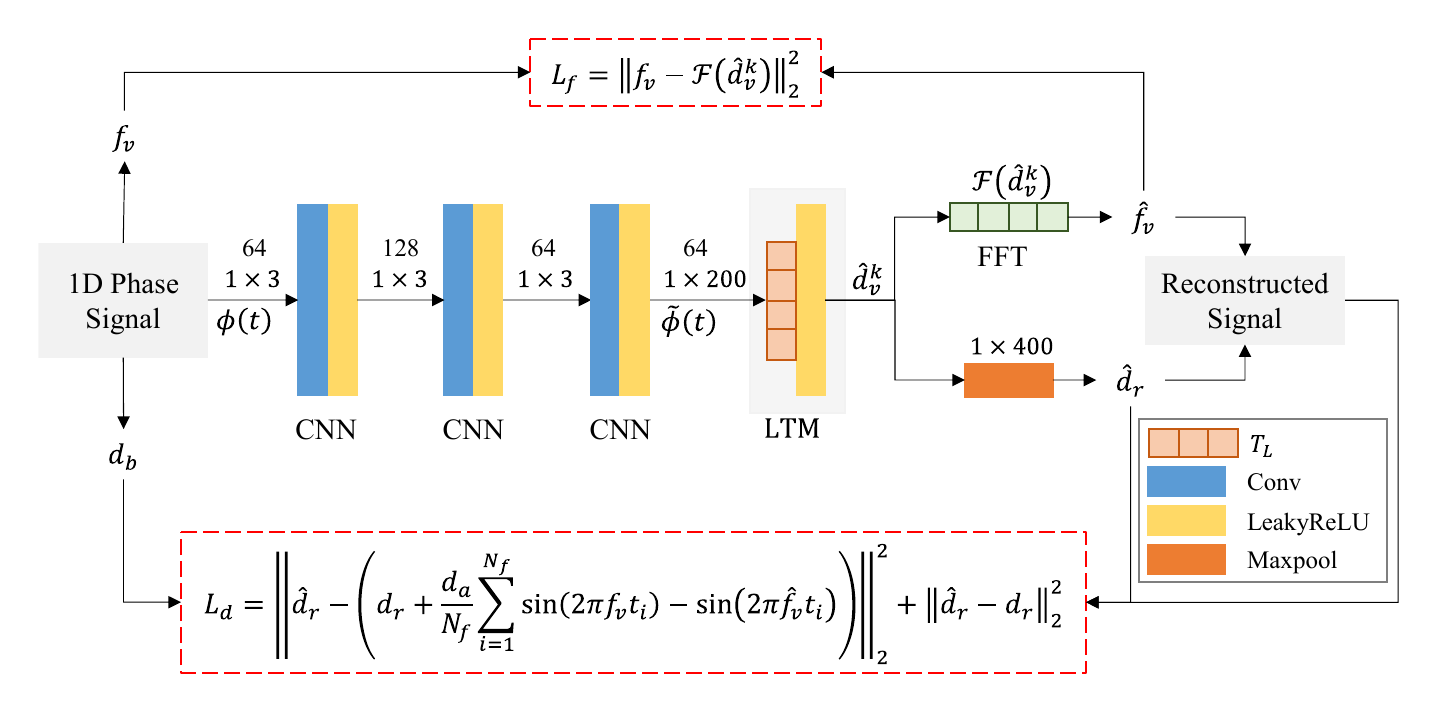}
    \caption{LTM-based Neural Network for mDD estimation.}
    \label{fig:ltmnn}
\end{figure}

In this section, we present the implementation details of the \ac{ltm} approach for \ac{isac}-based \ac{mdd} estimation. Additionally, we summarize the design mechanisms for network layers and loss functions, as shown in Fig.~\ref{fig:ltmnn}. 
The proposed \ac{ltm}-based interpretable network achieves the \ac{mdd} estimation and the clutter suppression using the input phase signal acquired by the \ac{bs} platform. Firstly, the input phase signal undergoes the feature transformation through a three-layer \ac{cnn} to produce the unwrapped phase $\boldsymbol{\widetilde{\phi}}(t)$. 
In particular, each \ac{cnn} layer employs a narrow-band convolution kernel with dimensions of $1\times 3$ to effectively capture local characteristics of the input signal with the LeakyReLU being the activation function. 
The narrow-band convolution ensures local continuity of phase information, while the LeakyReLU activation addresses potential truncation issues of phase values near zero. This strategy maintains global continuity and consistency throughout the phase-mapping process. 
As depicted in Fig.~\ref{fig:ltmnn}, the proposed \ac{cnn}-based architecture efficiently performs length-preserving transformations within the input phase, ultimately yielding accurate unwrapped phase information $\boldsymbol{\widetilde{\phi}}(t)$.

Subsequently, the unwrapped phase signal is processed through an \ac{ltm}-based signal decoupling procedure. Specifically, a single \ac{cnn} layer implements parameterized template matching, converting the phase signal into deformation displacement signals as described in (\ref{ts-match}) and (\ref{o}). Then, the learnable template $\boldsymbol{T}_L\in \mathbbm{R}^{1\times 200}$ with the stride of $1$ is obtained from the network to extract the match target signal features while mitigating the clutter signal. Moreover, we design a max pooling operation with kernel size $N_p=400$ (twice the template length) to further enhance and extract the matched \ac{mdd} signal and suppress unmatched clutter. {\color{black}Finally, the output of the \ac{ltm}-based network is the estimated displacement signal $\hat{d}_r$ and its corresponding vibration frequency $\hat{f}_v$.} Leveraging these estimated parameters, the reconstructed \ac{mdd} signal is calculated as
\begin{equation}
    \begin{aligned}
        \centering
        \boldsymbol{\hat{d}}_v = \hat{d}_r + \alpha \hat{d}_r\sin \left(2\pi\hat{f}_v\frac{l-1}{N}\right),~l=1,...,N.
    \end{aligned}
    \label{mdd_reconstruct}
\end{equation}

According to the proposed \ac{ltm}-based architecture, we define a loss function incorporating the estimated displacement, vibration frequency, and reconstructed \ac{mdd} signal, expressed as
\begin{equation}
    \begin{aligned}
        \centering
        L = \Vert \hat{d}_r - d_r \Vert_2^2 + \Vert f_v - \mathscr{F}\left(\boldsymbol{\hat{d}}^k_v\right)\Vert_2^2+L_d,
    \end{aligned}
    \label{loss}
\end{equation}
{\color{black}where the $L_r = \Vert \hat{d}_r - d_r \Vert_2^2$ denotes the estimated displacement loss, and $L_f =\Vert f_v - \mathscr{F}\left(\boldsymbol{\hat{d}}^k_v\right)\Vert_2^2$ represents the vibration frequency estimation loss}, with $\mathscr{F}\left(\boldsymbol{\hat{d}}^k_v\right)$ indicating the spectrum of the output \ac{mdd} signal. Additionally, the signal reconstruction loss $L_d$ can be calculated as
\begin{equation}
    \begin{aligned}
        \centering
        L_d = \Vert \hat{d}_r - \left(d_r + \frac{d_a}{N_f}\sum_{i=1}^{N_f}\sin \left(2\pi f_vt_i\right)-\sin \left(2\pi \hat{f}_vt_i\right)\right) \Vert_2^2.
    \end{aligned}
    \label{rec_loss}
\end{equation}
For convenience of analysis, we assume the \ac{mdd} signal remains the fixed vibration frequency within one sensing pulse duration $N_f$. 
Accordingly, the deformation displacement reconstruction loss can be simplified as
\begin{equation}
    \begin{aligned}
        \centering
        L_d = \Vert \hat{d}_r - \left(d_r + \frac{d_a}{N_f}N_f\left(\sin \left(2\pi f_vt_i\right)-\sin \left(2\pi \hat{f}_vt_i\right)\right)\right) \Vert_2^2.
    \end{aligned}
\end{equation}

\begin{figure}[tbp]
    \centering
    \includegraphics[width=0.95\linewidth]{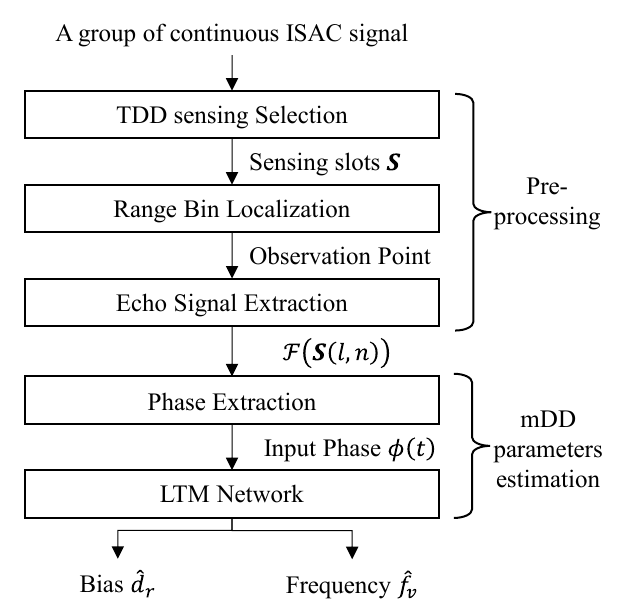}
    \caption{Deformation measurement flow chart.}
    \label{fig:workflow}
\end{figure}

To sum up, the proposed \ac{ltm}-based interpretable network effectively estimates displacement and vibration frequency from the phase signal acquired by the \ac{bs} platform. By integrating target-clutter feature extraction and an \ac{fft} operator into the network architecture, the proposed method reconstructs the \ac{mdd} signals at observation points, facilitating precise dynamic micro-deformation monitoring. {\color{black}Fig.~\ref{fig:workflow} illustrates the overall \ac{mdd} parameter estimation workflow based on the \ac{isac} platform, incorporating the proposed LTM-based interpretable network.}

	%----------------------------------------------------------------------------------------
	%	NUMERICAL EVALUATIONS
	%----------------------------------------------------------------------------------------

	\section{Numerical Evaluations}
	\label{sec:Sims}
	
In this section, we present numerical results to evaluate the performance of our proposed \ac{isac}-based \ac{mdm} systems and the proposed algorithms. 
Specifically, in section~\ref{subsec:setup}, we describe the simulation framework for the \ac{isac} system and the preparation of the \ac{mdm} training dataset. Subsequently, we present numerical results for the urban scenarios in section~\ref{subsec:converg}, aiming to assess the \ac{mdd} estimation accuracy and clutter suppression effectiveness.

\subsection{Simulation Setup}
\label{subsec:setup}

\textbf{ISAC-based mDM Simulation}:
{\color{black}The proposed \ac{isac} system configuration is shown in Table~\ref{tab:isacplat}, and the operation wavelength is $\lambda=0.0612$~m}. 
The transmit and receive antennas are adjacent in the XOZ plane. Simultaneously, all communication users and observation targets are situated within the far-field radiation region of the \ac{bs}, represented by the Cartesian coordinate system. 
\begin{table}[htbp]
\centering
\caption{The \ac{bs}-based \ac{isac} platform configuration}
\label{tab:isacplat}
\begin{tabularx}{0.99\columnwidth}{lc}
\toprule \toprule
Parameters & Value \\
\midrule
Center frequency               & 4.9~GHz \\
Bandwidth                      & 100~MHz \\
PRF                            & 1~kHz \\
Samples per pulse              & 200 \\
Position of the transmitter & (0, 0, 40.896)~m \\
Position of the receiver  & (0.003, 0.311, 46.396)~m \\
Azimuth angle of the target    & $8^{\circ}$ \\
Look-down angle of the target  & $27^{\circ}$ \\
\bottomrule \bottomrule
\end{tabularx}
\end{table}

\textbf{Dataset and Metrics}:
We conducted experiments on the numerical simulated dataset, the detailed dataset preparation is shown in Table~\ref{tab:trainset}. 
To improve the generalization of \ac{mdm} simulation data, \ac{awgn} is introduced to the \ac{bs} received signal during simulations, with the \ac{snr} varying from $0$~dB to $10$~dB in $1$~dB increments. 
We randomly select $30976$ samples, with $18586$ for training, $4646$ for validation, and $7744$ for testing. 
This approach guarantees the distinctiveness of the training and testing datasets. Notably, the samples fed into the \ac{ltm} network can be distinguished by at least one varying parameter involving bias, frequency, or \ac{snr}.
\begin{table}[htbp]
\centering
\caption{Training dataset preparation}
\label{tab:trainset}
\begin{tabularx}{0.99\columnwidth}{lc}
\toprule \toprule
Parameters & Value \\
\midrule
Samples per frame             & 1000 \\
Duration per frame            & 1~s \\
Training set                  & 23232 (75\%) \\
Testing set                   & 7744 (25\%) \\
Bias interval of the mDD       & 0--25~mm \\
Frequency interval of the mDD  & 5--15~Hz \\
SNR interval                  & 0--10~dB \\
Training epoch                & 40 \\
\bottomrule \bottomrule
\end{tabularx}
\end{table}

\subsection{Performance Analysis}
\label{subsec:converg}
\begin{figure*}[htbp]
	\centering  
	\vspace{-0.35cm} %设置与上面正文的距离
	\subfigtopskip=2pt %设置子图与上面正文或别的内容的距离
	\subfigbottomskip=2pt %设置第二行子图与第一行子图的距离，即下面的头与上面的脚的距离
	\subfigcapskip=-5pt %设置子图与子标题之间的距离
	\subfigure[]{
		\label{fig:sim_loss}
		\includegraphics[width=0.95\linewidth]{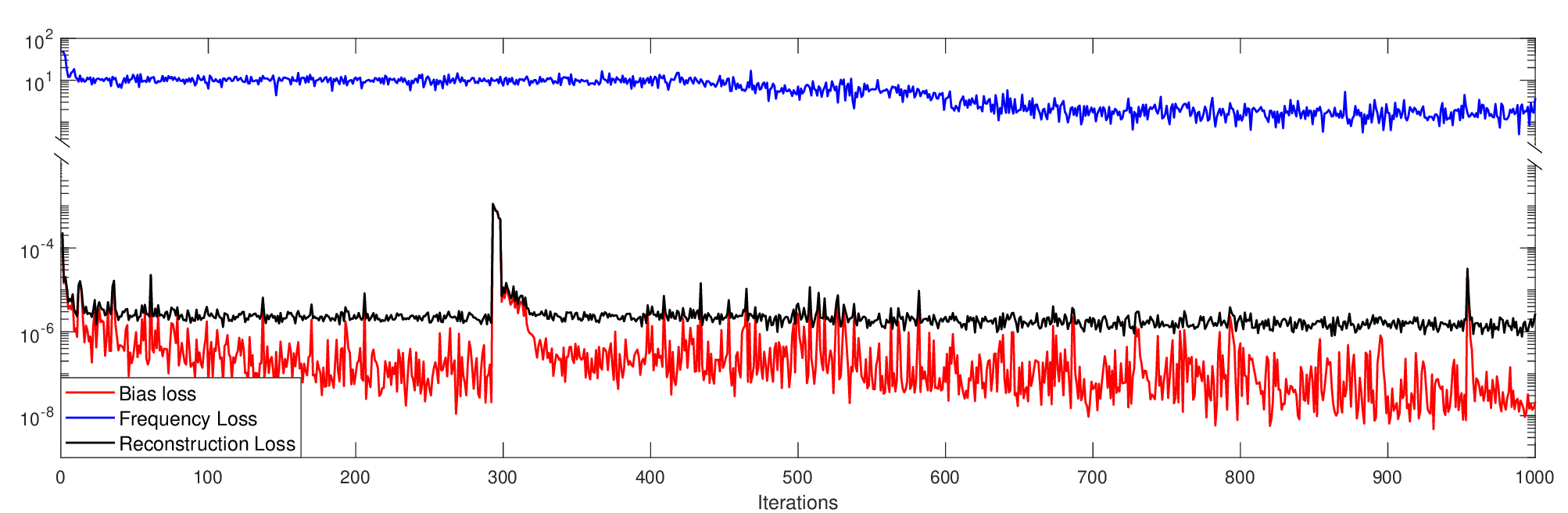}}
  
	\subfigure[]{
		\label{fig:sim_esti}
		\includegraphics[width=0.95\linewidth]{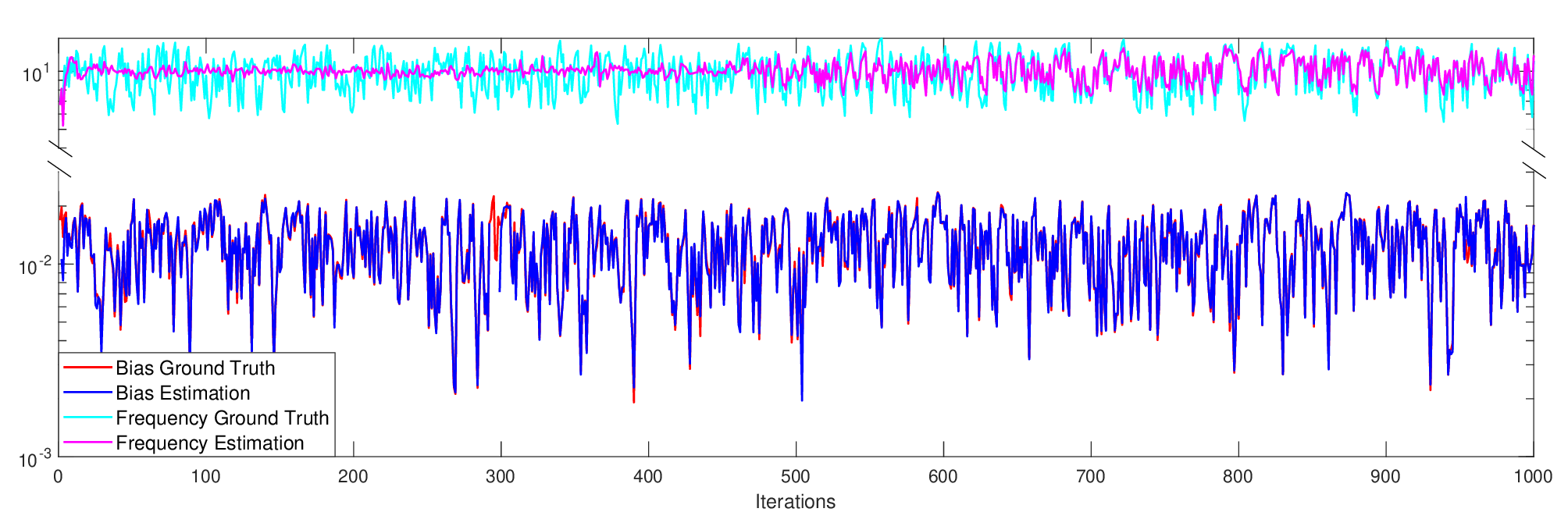}}
	\caption{Illustration of LTM Convergence: (a) {\color{black} Convergence of the LTM neural network, showing the progression of bias loss, frequency loss, and reconstruction loss over mini-batch training iterations}. (b) Reconstruction fidelity, providing the evolution of bias and frequency reconstruction results {\color{black}during mini-batch network training}.}
	\label{fig:convergence}
\end{figure*}

First, we analyze the convergence performance of the proposed \ac{ltm}-based neural network, where the training dataset is prepared as Table~\ref{tab:trainset}. In particular, by applying the \ac{ltm} method, the convergence of the \ac{ltm}-based interpretable network is illustrated in Fig.~\ref{fig:convergence}. Specifically, the loss function of the \ac{ltm} network consists of three components: displacement bias estimation loss, frequency estimation loss, and \ac{mdd} signal reconstruction loss. 
As shown in Fig.~\ref{fig:sim_loss}, the displacement bias loss converges rapidly and stabilizes within one epoch ($25$ iterations), whereas the frequency estimation loss requires over $25$ epochs (approximately $600$ iterations) to fully converge. Furthermore, since the displacement bias primarily influences the \ac{mdd} signal, the reconstruction loss converges almost simultaneously with the displacement bias loss.  

To further verify the reconstruction accuracy of the proposed \ac{ltm}-based network, we evaluate the estimated \ac{mdd} signals against the ground truth during the training process, as depicted in Fig.~\ref{fig:sim_esti}. Specifically, we compare the displacement estimated by the \ac{mdd} network with its corresponding ground truth and analyze the evolution of frequency estimation accuracy with increasing training iterations. From Fig.~\ref{fig:sim_esti}, it can be seen that the estimated displacement rapidly aligns with the ground truth during early training stages, whereas the accuracy of frequency estimation progressively improves as training continues. These results indicate that the \ac{ltm}-based network initially achieves robust displacement reconstruction and subsequently refines its capability to learn the frequency characteristics of the \ac{mdd} signal. Motivated by these training outcomes, the following experiments employ the trained \ac{ltm} network to evaluate its performance on \ac{mdd} data processing in various \ac{isac}-\ac{mdm} application scenarios.

\subsubsection{Simulation Result}
\label{multi-user single target}

In this section, we evaluate the generalization performance of the \ac{ltm}-based interpretable neural network using a test dataset prepared according to the parameters listed in Table~\ref{tab:trainset}. As detailed previously in Section~\ref{subsec:setup} regarding the experimental parameters of the \ac{isac}-\ac{mdm} simulation platform, each training sample differs in one or more key attributes, such as displacement bias, vibration frequency, or \ac{snr}. Consequently, in the generalization experiments, the parameters of all test samples are selected to differ from those in the training dataset. Specifically, none of the \ac{mdd} signals in the test dataset share identical parameter configurations with any sample included in the training set.

\begin{figure}
	\centering  
	\vspace{-0.35cm} %设置与上面正文的距离
	\subfigtopskip=2pt %设置子图与上面正文或别的内容的距离
	\subfigbottomskip=2pt %设置第二行子图与第一行子图的距离，即下面的头与上面的脚的距离
	\subfigcapskip=-5pt %设置子图与子标题之间的距离
	\subfigure[]{
		\label{fig:sim_bias}
		\includegraphics[width=0.95\linewidth]{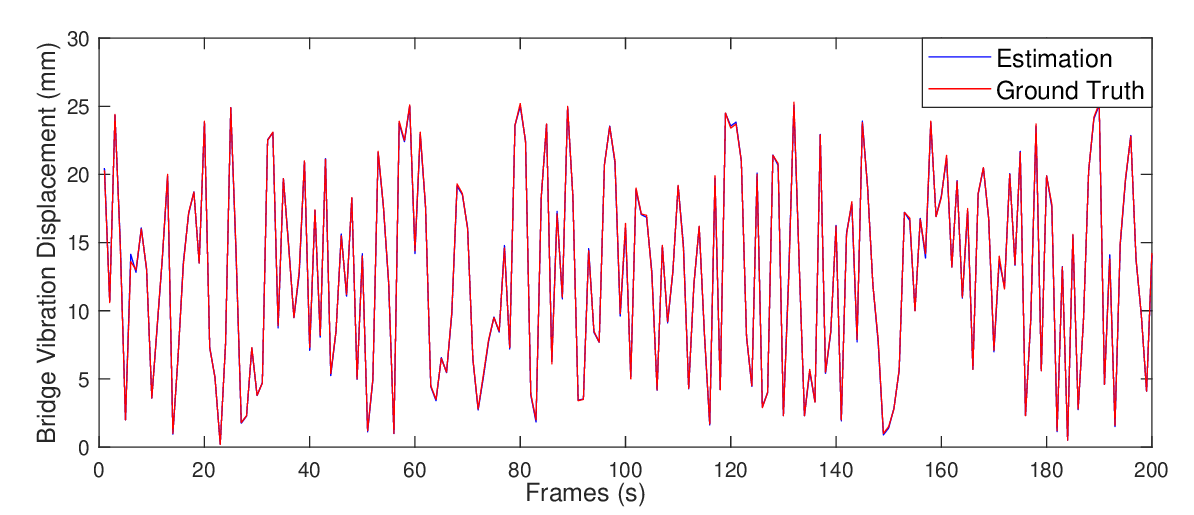}}
  
	\subfigure[]{
		\label{fig:sim_freq}
		\includegraphics[width=0.95\linewidth]{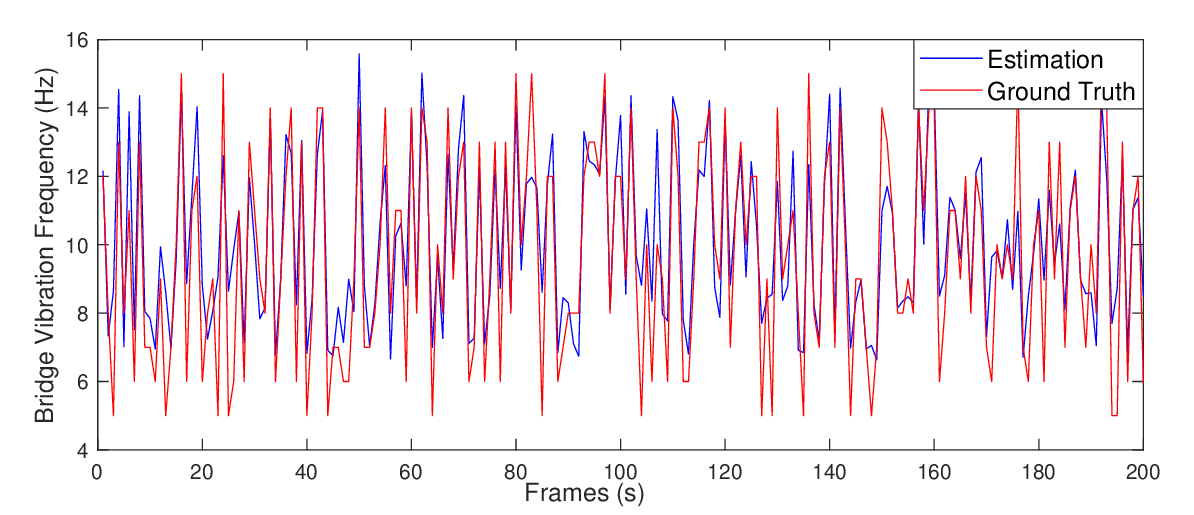}}
	\caption{mDD estimation results of the LTM network. (a) Bias estimation results of the mDD signal, demonstrating the accuracy of the estimated displacement bias. (b) Frequency estimation results of the mDD signal, showing the precision of the estimated vibration frequency.}
	\label{fig:simexp}
\end{figure}

Based on the previously defined principle for sample selection in the test dataset, the generalization performance of the proposed \ac{ltm}-based interpretable neural network is presented in Fig.~\ref{fig:simexp}. Specifically, Fig.~\ref{fig:sim_bias} demonstrates the capability of the \ac{ltm}-based network in estimating displacement bias across all $7744$ test samples. Numerical results show that the proposed method accurately and efficiently estimates displacement within the entire dynamic range tested ($0$ to $25$ mm). Additionally, Fig.~\ref{fig:sim_freq} evaluates the frequency estimation performance, indicating accurate predictions for the majority of samples, with only a small portion exhibiting estimation errors of approximately $2$ Hz. The effective dynamic range for frequency estimation extends from $5$ to $15$ Hz.
Considering the overall \ac{ltm}-based network performance, we selected $40$ epochs as the convergence stage to achieve an optimal balance between displacement bias and frequency estimation accuracy. Under this setting, the proposed method ensures precise displacement estimation, while minor frequency estimation deviations have negligible impact on the overall accuracy of the reconstructed \ac{mdd} signals.

\subsubsection{BS Simulation}
\label{multi-user multi-target}

The analyses presented in previous subsections evaluated the \ac{ltm}-based interpretable neural network’s performance under scenarios involving various \ac{awgn}. 
In this section, we extend our investigation to the performance of \ac{mdd} signal extraction and clutter suppression, specifically focusing on clutter interference characterized by periodic vibration signals. 
Our primary goal is to further validate the parameter estimation accuracy and reconstruction capability of the proposed network for the \ac{mdd} signal. 
To this end, we generate simulated phase data representing deformation displacement signals that include the \ac{mdd} signal, fixed-frequency clutter, and random noise components.

\begin{table}[htbp]
\centering
\caption{The \ac{mdd} signal simulation setup}
\label{tab:mddsim}
\begin{tabularx}{0.99\columnwidth}{lc}
\toprule \toprule
mDD & Number of times \\
\midrule
Random samples in $[-0.15, 0.15]$~mm & 2000 \\
Random samples in $[0.5, 1]$~mm      & 30 \\
Random samples in $[1, 2]$~mm        & 5 \\
Random samples in $[2, 3]$~mm        & 2 \\
\bottomrule \bottomrule
\end{tabularx}
\end{table}

Concerning the design of clutter signals, a single-frequency vibration interference at $0.41$~Hz is generated as the dominant disturbance source throughout the observation period. Additionally, to mimic the practical noise conditions, \ac{awgn} is incorporated by adding the random fluctuations with displacement bias of $\pm 0.15$~mm onto the original deformation displacement signal. Furthermore, to accurately simulate practical deformation scenarios, a series of deformation signals with displacement bias exceeding $0.5$~mm are generated. The detailed simulation parameters are provided in Table~\ref{tab:mddsim}.

\begin{figure}
	\centering  
	\vspace{-0.35cm} %设置与上面正文的距离
	\subfigtopskip=2pt %设置子图与上面正文或别的内容的距离
	\subfigbottomskip=2pt %设置第二行子图与第一行子图的距离，即下面的头与上面的脚的距离
	\subfigcapskip=-5pt %设置子图与子标题之间的距离
	\subfigure[]{
		\label{fig:BVM_real}
		\includegraphics[width=0.95\linewidth]{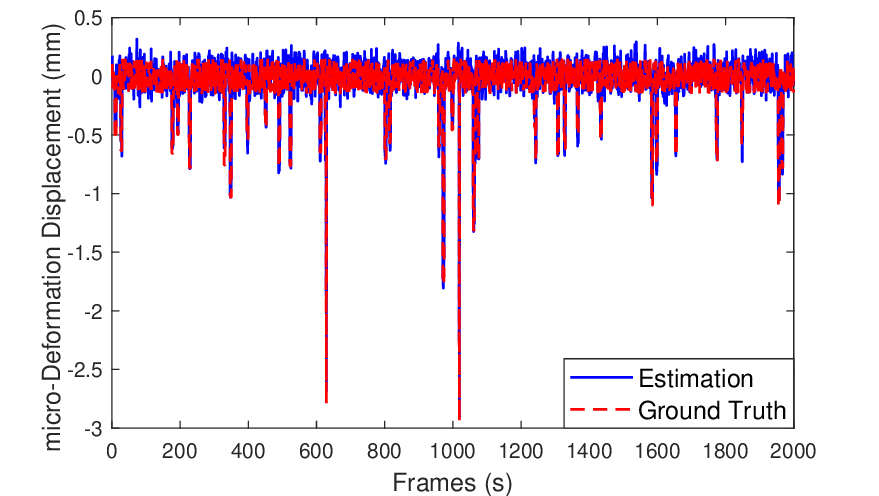}}
  
	\subfigure[]{
		\label{fig:BVM_real_fft}
		\includegraphics[width=0.95\linewidth]{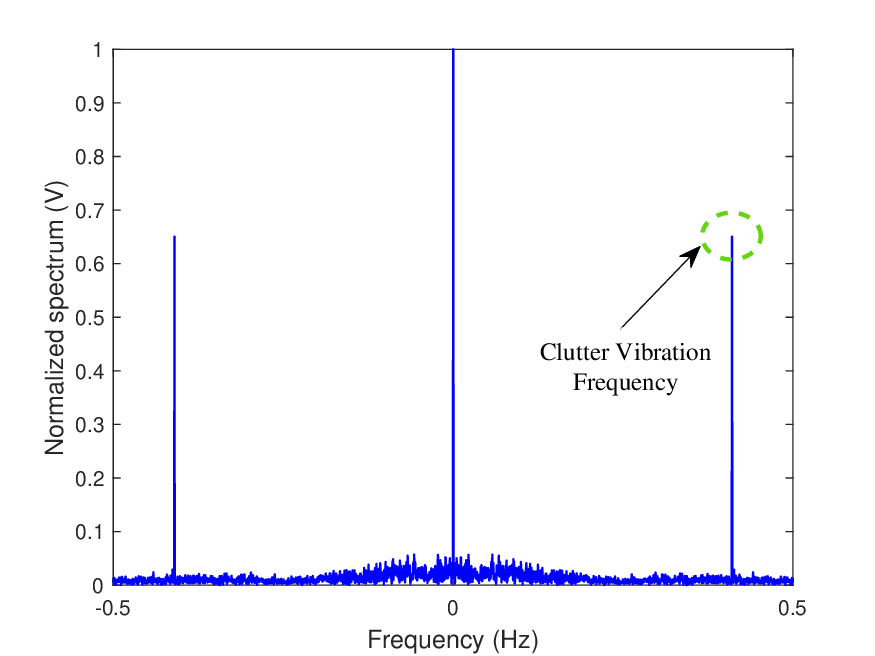}}
	\caption{Simulation results of the LTM network. (a) The mDD results obtained from the LTM network. (b) Spectrum of mDD results using the LTM network.}
	\label{fig:realexp}
\end{figure}

Based on the \ac{isac}-\ac{mdm} configuration, the generated phase signals extracted from the \ac{bs} platform are sequentially fed into the pre-trained \ac{ltm} network, with each signal frame consisting of $1000$ sampling points. Fig.~\ref{fig:BVM_real} illustrates the estimated displacement bias of the \ac{mdd} signal after eliminating both fixed-frequency clutter interference and random background fluctuations. The experimental results demonstrate that the \ac{ltm}-based network achieves high-precision \ac{mdd} estimation, effectively suppressing both random noise and periodic vibration interference.
Furthermore, Fig.~\ref{fig:BVM_real_fft} presents a spectral analysis of the reconstructed \ac{mdd} signal, clearly demonstrating the network’s capability to separate the deformation signal from background clutter. Specifically, the zero-frequency component corresponds to the displacement bias due to \ac{mdd} during the observation period, while the component at $0.41$~Hz verifies the presence of fixed-frequency clutter in the original deformation signal.
Numerical results in Fig.~\ref{fig:realexp} indicate that the proposed \ac{ltm} network effectively operates in cluttered environments. Particularly, under more complex interference scenarios, the proposed method exhibits strong clutter suppression performance and precise estimation of signal parameters.

\begin{figure}
	\centering  
	\vspace{-0.35cm} %设置与上面正文的距离
	\subfigtopskip=2pt %设置子图与上面正文或别的内容的距离
	\subfigbottomskip=2pt %设置第二行子图与第一行子图的距离，即下面的头与上面的脚的距离
	\subfigcapskip=-5pt %设置子图与子标题之间的距离
	\subfigure[]{
		\label{fig:BVM_real_static}
		\includegraphics[width=0.95\linewidth]{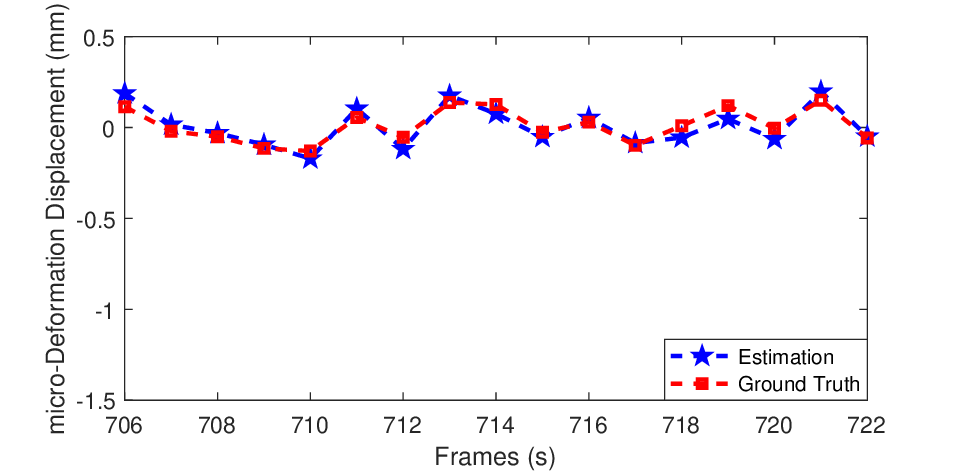}}
  
	\subfigure[]{
		\label{fig:BVM_real_vib}
		\includegraphics[width=0.95\linewidth]{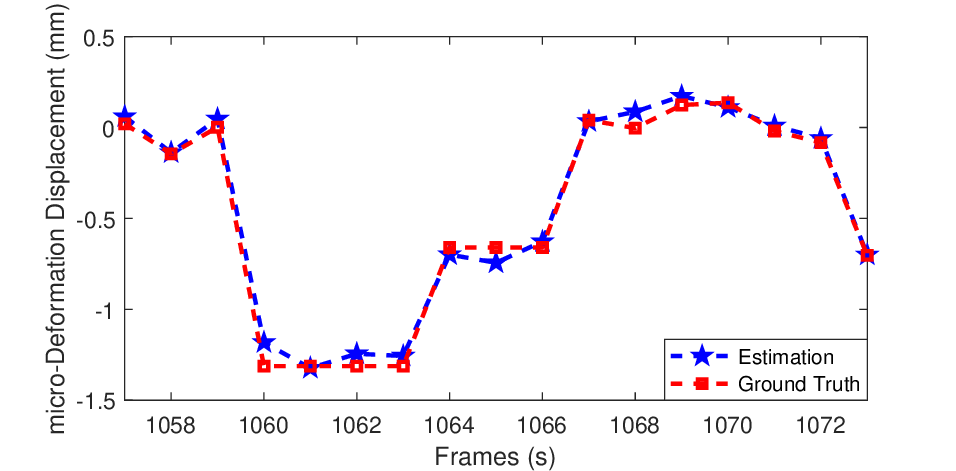}}
	\caption{mDD Estimation Results Based on the BS Platform (a) mDD estimation results during non-vehicle intervals. (b) mDD estimation results during instances of overloading vehicles.}
	\label{fig:realexpin}
\end{figure}

To further validate the predictive capability of the proposed \ac{ltm}-based network, Fig.~\ref{fig:realexpin} presents the \ac{mdd} estimation results under clutter interference. The experiment considers both static and deformation states at the observation point. For consistency, the observation duration is set to $17$ seconds for each condition.
As shown in Fig.~\ref{fig:BVM_real_static} and Fig.~\ref{fig:BVM_real_vib}, the pre-trained network accurately estimates micro-deformation displacements in both static and deformation scenarios. During non-vehicle intervals, the \ac{ltm}-based network effectively suppresses single-frequency vibration clutter, maintaining displacement fluctuations within $\pm 0.15$~mm. In contrast, during overloading events where infrastructure displacements exceed $1$~mm, the network continuously tracks large deformation events lasting up to $7$ seconds. For example, in Fig.~\ref{fig:BVM_real_vib}, the network captures a displacement of approximately $1.5$ mm between $1060$–$1063$ seconds, followed by $0.7$ mm deformations from $1064$–$1066$ seconds, and again after a static period between $1067$–$1072$ seconds.
These results confirm that the proposed \ac{ltm}-based interpretable neural network enables accurate, continuous, and long-term monitoring of the \ac{mdd} signals. Furthermore, it ensures reliable tracking in both static and deformation states, demonstrating high efficiency and robustness in practical \ac{isac}-based micro-deformation monitoring scenarios.

\subsubsection{BS Experiments}
\label{bs experiments}
\begin{figure}[htbp]
    \centering
    \includegraphics[width=0.95\linewidth]{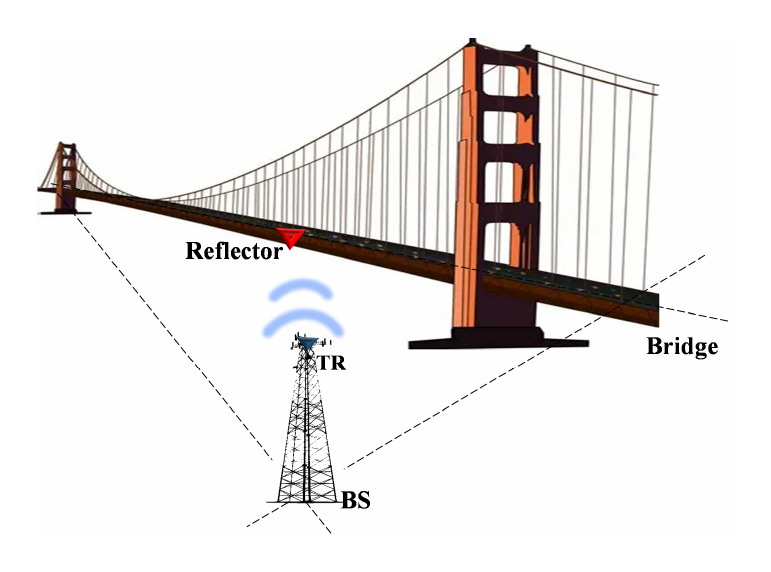}
    \caption{{\color{black}Sketch of the bridge vibration deformation measurement setup using the BS platform.}}
    \label{fig:bsexp}
\end{figure}

{\color{black}To evaluate the effectiveness of the proposed method in practical scenarios, we conduct experiments on a real \ac{bs} platform and measure the \ac{mdd} of the Nanjing Qixiashan Yangtze River Bridge. The experiment focuses on bridge vibration deformation, which serves as a representative mDD task. The measurement geometry of the BS setup is illustrated in Fig.~\ref{fig:bsexp}, while the operating parameters are summarized in Table~\ref{tab:isacplat}. In this configuration, the BS transmits FMCW signals in TDD mode, and the received echoes are processed to achieve high-accuracy bridge vibration displacement estimation.}

{\color{black}
\textbf{Comparison}:
To comprehensively evaluate the effectiveness of the proposed LTM in addressing the mDD problem, we perform comparative experiments with several representative regression-based deep learning algorithms, a fixed-template filtering approach (denoted as SP), and ablation studies of LTM under different loss functions. The deep learning baselines include three widely used architectures for processing temporal radar signals: CNN \cite{lin2023accurate}, MLP \cite{nguyen2025application}, and LSTM \cite{shan2025damage}.
In addition, we conduct ablation tests to examine the impact of different loss components. As defined in (17), the overall loss consists of three parts: $L_r$, $L_f$, and $L_d$. To assess their contributions, we remove each component individually and denote the resulting variants as LTM-$L_r$ (without $L_r$), LTM-$L_f$ (without $L_f$), and LTM-$L_d$ (without $L_d$).
All methods are trained on the same dataset to ensure fairness, and their performance is validated on real measurement data. The evaluation considers both the estimated deformation displacement and the frequency of large deformation events, where large deformations are defined as those exceeding $1$ mm and $2$ mm. This evaluation framework provides a comprehensive assessment of the reliability and robustness of different methods in practical applications.}

\begin{figure}
	\centering  
	\vspace{-0.35cm} %设置与上面正文的距离
	\subfigtopskip=2pt %设置子图与上面正文或别的内容的距离
	\subfigbottomskip=2pt %设置第二行子图与第一行子图的距离，即下面的头与上面的脚的距离
	\subfigcapskip=-5pt %设置子图与子标题之间的距离
	\subfigure[]{
		\label{fig:BS_real}
		\includegraphics[width=0.95\linewidth]{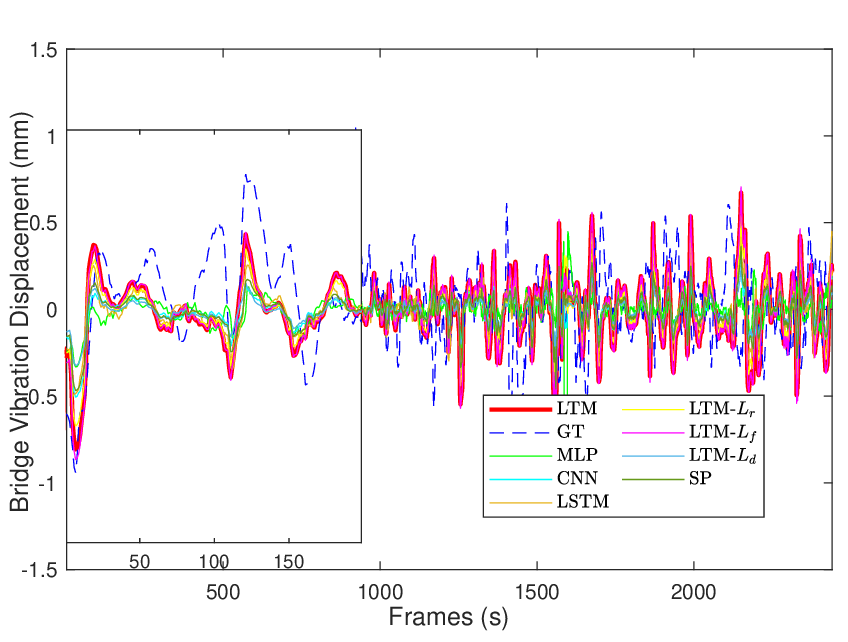}}
  
	\subfigure[]{
		\label{fig:BS_real_fft}
		\includegraphics[width=0.95\linewidth]{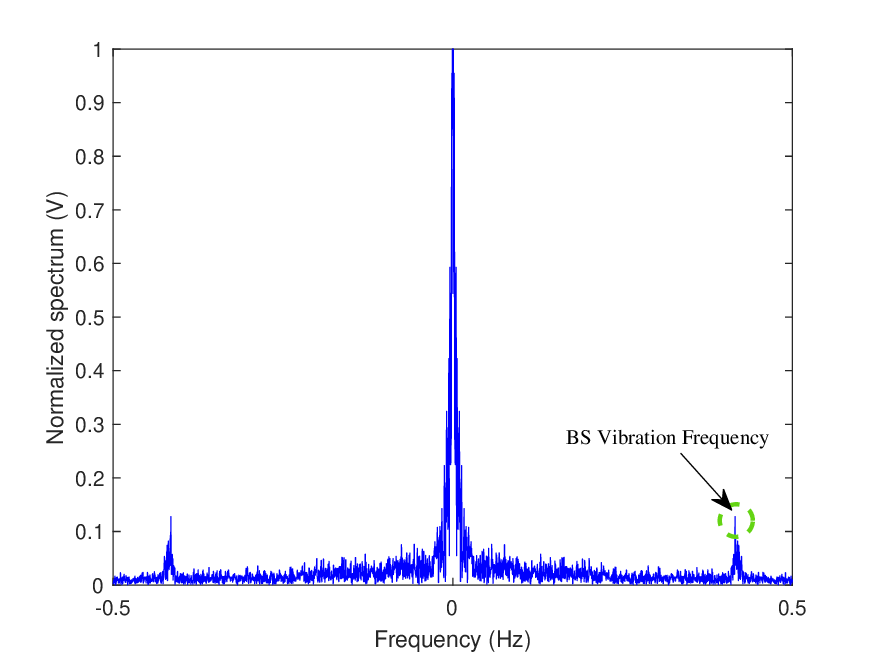}}
	\caption{Bridge vibration displacement estimation results of the BS platform. (a) Comparison of the proposed LTM, its ablation variants, and other baseline methods. (b) Spectrum of Bridge vibration results.}
	\label{fig:BSrealexp}
\end{figure}

\begin{figure}
	\centering  
	\vspace{-0.35cm} %设置与上面正文的距离
	\subfigtopskip=2pt %设置子图与上面正文或别的内容的距离
	\subfigbottomskip=2pt %设置第二行子图与第一行子图的距离，即下面的头与上面的脚的距离
	\subfigcapskip=-5pt %设置子图与子标题之间的距离
	\subfigure[]{
		\label{fig:BS_real_static}
		\includegraphics[width=0.95\linewidth]{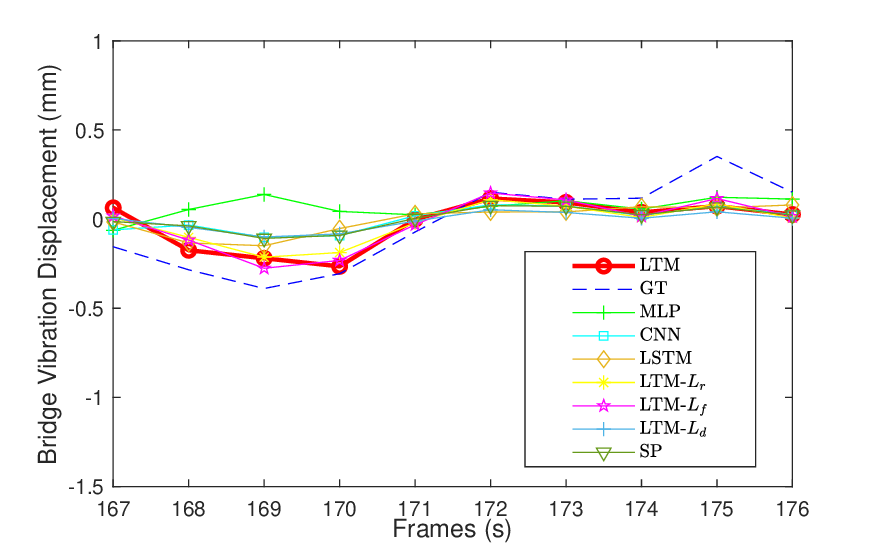}}
  
	\subfigure[]{
		\label{fig:BS_real_vib}
		\includegraphics[width=0.95\linewidth]{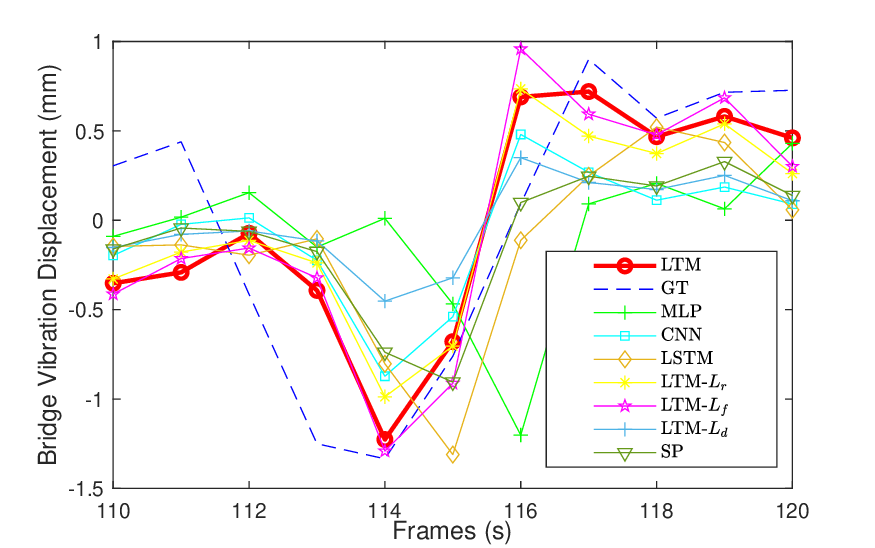}}
	\caption{Bridge vibration estimation results on the BS platform. (a) Estimated displacements during non-vehicle intervals. (b) Estimated displacements during overloaded-vehicle passages.}
	\label{fig:BSrealexpin}
\end{figure}

{\color{black}The calibrated results of bridge vibration deformation estimation are presented in Fig.~\ref{fig:BSrealexp}. In Fig.~\ref{fig:BS_real}, we evaluate the performance of the estimated displacements obtained by the proposed and baseline algorithms over a continuous observation period of $2440$~s. All sequences are smoothed using a uniform sliding-average filter to highlight long-term trends. To provide a clearer view of the displacement details, the first $150$~s are enlarged. These results clarify that the proposed LTM and its variants yields the closest match to the ground truth, particularly in capturing the deformation trend, followed by the LSTM-based method. In contrast, the MLP-based method exhibits the lowest accuracy.
Fig.~\ref{fig:BSrealexp} further shows the spectral characteristics of the bridge vibration displacements, revealing a coupling effect with the inherent vibration of the BS. Despite this coupling, the proposed LTM maintains accurate deformation estimation, demonstrating strong robustness under real measurement conditions.

To investigate the impact of each component in the loss function, we present the estimated displacements under different bridge states in Fig.~\ref{fig:BSrealexpin}, along with the corresponding quantitative results in Table~\ref{tab:mddexp}. 
Fig.~\ref{fig:BS_real_static} and Fig.~\ref{fig:BS_real_vib} illustrate the estimated displacements of the static state and the large deformation state of the bridge, respectively. It's obvious that all algorithms perform well in the static state. 
However, under large deformation conditions, the proposed algorithm significantly outperforms both its ablated variants and the baseline methods. 
In particular, the LTM-$L_f$ variant achieves performance comparable to the full LTM model and is superior to LTM-$L_d$. The LTM-$L_r$ variant exhibits the lowest accuracy among the three.
These results indicate that the loss term $L_r$ plays a crucial role in the mDD task, while the $L_d$ term, which corresponds to the vibration reconstruction loss, also contributes positively to network training. In contrast, the $L_f$ term has a relatively minor effect on the performance of the LTM network.

To accurately assess the performance of the proposed algorithm, its variants, and baseline algorithms, the number of large deformation events under two error tolerances is detailed in Table \ref{tab:mddexp}.
Specifically, we count the number of estimated displacements exceeding $1$~mm and $2$~mm, within error margins of $\delta_1 = 0.1$ and $\delta_2 = 0.2$, corresponding to prediction ranges of $1 \pm \delta_1$ mm and $2 \pm \delta_2$ mm, respectively.
According to the ground truth, there are $37$ events with displacements exceeding $1$~mm and $3$ events exceeding $2$~mm. The proposed LTM and the LTM-$L_f$ variant achieve the most accurate estimates of these events, closely aligning with the ground truth. 
In contrast, the baseline algorithms fail to capture nearly half of the large deformation occurrences, limiting their suitability for the mDD measurement using the BS platform.
These results demonstrate that the proposed LTM, along with its tailored loss function, outperforms other approaches and effectively addresses the ISAC-based mDD problem.  }

\begin{table}[htbp]
\centering
\caption{Bridge vibration displacement measurements from real-world experiments using the BS platform}
\label{tab:mddexp}
\begin{tabularx}{0.96\columnwidth}{lcccc}
\toprule \toprule
Methods \ \  \ &  \multicolumn{2}{c}{\makecell{Displacement \\ $>1$mm}}   &
          \multicolumn{2}{c}{\makecell{Displacement \\ $>2$mm}} \\
\cmidrule(lr){2-3} \cmidrule(lr){4-5} 
        & $\delta_1=0.1$ & $\delta_2=0.2$ & $\delta_1=0.1$ & $\delta_2=0.2$ \\
\midrule
LTM     & \textbf{27} & \textbf{34} & \textbf{4} & \textbf{4} \\
LTM-$L_r$  & 17 & 28 & 2 & 2 \\
LTM-$L_f$  & \textbf{35} & \textbf{41} & \textbf{4} & \textbf{4} \\
LTM-$L_d$  & 2  & 2  & 0 & 0 \\
CNN     & 10 & 14 & 1 & 1 \\
MLP     & 6  & 10 & 1 & 1 \\
LSTM    & 18 & 23 & 2 & \textbf{3} \\
SP      & 8  & 10 & 0 & 1 \\
\midrule
GT      & \multicolumn{2}{c}{\textbf{37}} & \multicolumn{2}{c}{\textbf{3}} \\
\bottomrule \bottomrule
\end{tabularx}
\end{table}

{\color{black}In summary, the proposed LTM-based interpretable neural network demonstrates high accuracy and robustness in ISAC-based mDD tasks. It consistently outperforms baseline algorithms and ablated variants in estimating both the deformation trend and large displacement events.
Furthermore, the LTM network maintains reliable performance under real-world measurement noise and structural vibration coupling, underscoring its practical potential for infrastructure deformation monitoring using ISAC systems.}

% \subsection{Discussion}
% \label{discussion}

	%----------------------------------------------------------------------------------------
	%	CONCLUSIONS
	%----------------------------------------------------------------------------------------
	\section{Conclusions}
	\label{sec:Conclusions}

In this paper, we introduced an \ac{isac}-based \ac{mdm} system designed for high-precision \ac{mdd} estimation in urban scenarios around \ac{bs}.
{\color{black}The shared hardware platform introduces sensing challenges such as low resolution and clutter interference, which pose significant challenges to accurate mDD estimation.}
To address these challenges, we described the environment model as a combination of \ac{bs} vibration and \ac{awgn}. 
We then formulated the \ac{mdd} reconstruction problem as a sensing signal enhancement task to facilitate phase unwrapping and \ac{mdd} signal decoupling.
Specifically, we proposed an \ac{ltm} approach that integrates a CNN-based phase unwrapping technique with a customized \ac{ltm} network architecture. 
{\color{black}Comprehensive evaluations were conducted under both simulation and real-world measurement settings. 
The proposed LTM network consistently outperformed baseline algorithms and its own ablated variants in estimating deformation trends and identifying large displacement events.} 
This work highlights the potential of \ac{isac}-based systems in infrastructure monitoring and paved the way for further advancements in \ac{isac} technology.
{\color{black}Moreover, the proposed \ac{ltm} framework provides a flexible solution for sensing tasks and can be extended to other scenarios involving weak periodic signal extractions.}

	\bibliographystyle{IEEEtran}
	\bibliography{IEEEabrv,main}

\end{document}